\documentclass[%
 reprint,
 amsmath,amssymb,
 aps,
 noeprint,
 superscriptaddress,
]{revtex4-2}

\usepackage{eqnarray}
\usepackage{xcolor}
\usepackage{graphicx}
\usepackage{dcolumn}
\usepackage{bm}
\usepackage{hyperref}
\usepackage{siunitx}

\begin{document}
\title{Distribution of non-Gaussian states in a deployed telecommunication fiber channel}

\newcommand{\DTU}{Center for Macroscopic Quantum States (bigQ), Department of Physics, Technical University of Denmark, Fysikvej, 2800 Kgs.\ Lyngby, Denmark}
\newcommand{\NICT}{Advanced ICT Research Institute, National Institute of Information and Communications Technology, 588-2 Iwaoka, Nishi-ku, Kobe 651-2492, Japan}
\newcommand{\NKT}{NKT Photonics A/S, Blokken 84, Birkerød DK-3460, Denmark}
\newcommand{\Kobe}{Graduate School of Engineering, Kobe University, 1-1 Rokkodai-cho, Nada, Kobe 657-0013, Japan}
\newcommand{\Tianjin}{College of Precision Instrument and Opto-Electronics Engineering, Tianjin University, Tianjin 300072, China}

\author{Casper A. Breum}
\altaffiliation[Present address: ]{\NKT}
\affiliation{\DTU}
\author{Xueshi Guo}
\altaffiliation[Present address: ]{\Tianjin}
\affiliation{\DTU}
\author{Mikkel V. Larsen}
\affiliation{\DTU}
\author{Shigehito Miki}
\affiliation{\NICT}
\author{Hirotaka Terai}
\affiliation{\NICT}
\author{Ulrik L. Andersen}
\affiliation{\DTU}
\author{Jonas S. Neergaard-Nielsen}
\email{jsne@fysik.dtu.dk}
\affiliation{\DTU}

\begin{abstract}
Optical non-Gaussian states hold great promise as a pivotal resource for advanced optical quantum information processing and fault-tolerant long-distance quantum communication. Establishing their faithful transmission in a real-world communication channel, therefore, marks an important milestone. In this study, we experimentally demonstrate the distribution of such non-Gaussian states in a functioning telecommunication channel that connects separate buildings within the DTU campus premises. We send photon-subtracted squeezed states, exhibiting pronounced Wigner negativity, through \SI{300}{m} of deployed optical fibers to a distant building. Using quantum homodyne tomography, we fully characterize the states upon arrival. Our results show the survival of the Wigner function negativity and coherence after transmission when correcting for detection losses, verifying the phase stability and low loss of the established link. 
This achievement not only validates the practical feasibility of distributing non-Gaussian states in real-world settings, but also provides an exciting impetus towards realizing fully coherent quantum networks for high-dimensional, continuous-variable quantum information processing.
\end{abstract}

\maketitle

\section{Introduction}
While most quantum information protocols to date are based on qubits defined by discrete energy levels of various physical systems, there is a rapidly growing interest in encoding them as  quantum states in the infinite-dimensional Hilbert space of bosonic harmonic oscillator modes. Examples of these include the Schr\"odinger cat codes and the Gottesman-Kitaev-Preskill (GKP) codes \cite{Gottesman2001,Ralph2003}. The allure of these bosonic codes lies in their inherent error-correcting capabilities \cite{Gottesman2001,Albert2018,Grimsmo2020} which in turn nourish promising ideas for fault-tolerant quantum computing \cite{Menicucci2014,Larsen2021,Tzitrin2021} and quantum communication \cite{Fukui2021,Rozpedek2021,Wu2022}. Numerous strides have been made on the experimental front with successful realizations of small cat states \cite{Ourjoumtsev2006,Neergaard-Nielsen2006} and cluster states suitable for GKP quantum computing in the optical regime \cite{Larsen2019,Asavanant2019}, as well as GKP state generation in trapped ions, superconducting circuits and optical modes \cite{Fluhmann2019,Campagne-Ibarcq2020,larsen_integrated_2025} and demonstrations of error correction beyond the break-even point using both cat states \cite{Ofek2016} and GKP states \cite{Sivak2023} in the microwave regime. 

One of the next frontiers is to leverage the power of these high-dimensional, non-Gaussian bosonic codes for quantum networking. The vision is to enable spatially separated nodes in a network to transmit or share non-Gaussian entangled codes. Such a quantum network would not only provide a medium for the coherent transfer of quantum information between quantum computer modules, but could also pave the way for communication with unconditional security over extreme distances when quantum error-correction is applied to the bosonic codes \cite{Albert2018,Fukui2021,Rozpedek2021,Wu2022,Hastrup2022}.
While the coherent transmission of single photon states and squeezed states in a hostile fiber or free-space environment have been demonstrated multiple times \cite{Yin2017,Wengerowsky2020,Neumann2022,Gyger2022,Peuntinger2014,Suleiman2022}, the transmission of non-Gaussian bosonic codes in an in-field, real-life environment remains unexplored territory. This is an essential first step towards the realization of a fully coherent quantum network for high-dimensional bosonic, continuous-variable codes.

\begin{figure*}
    \centering
    \includegraphics[width=.75\linewidth]{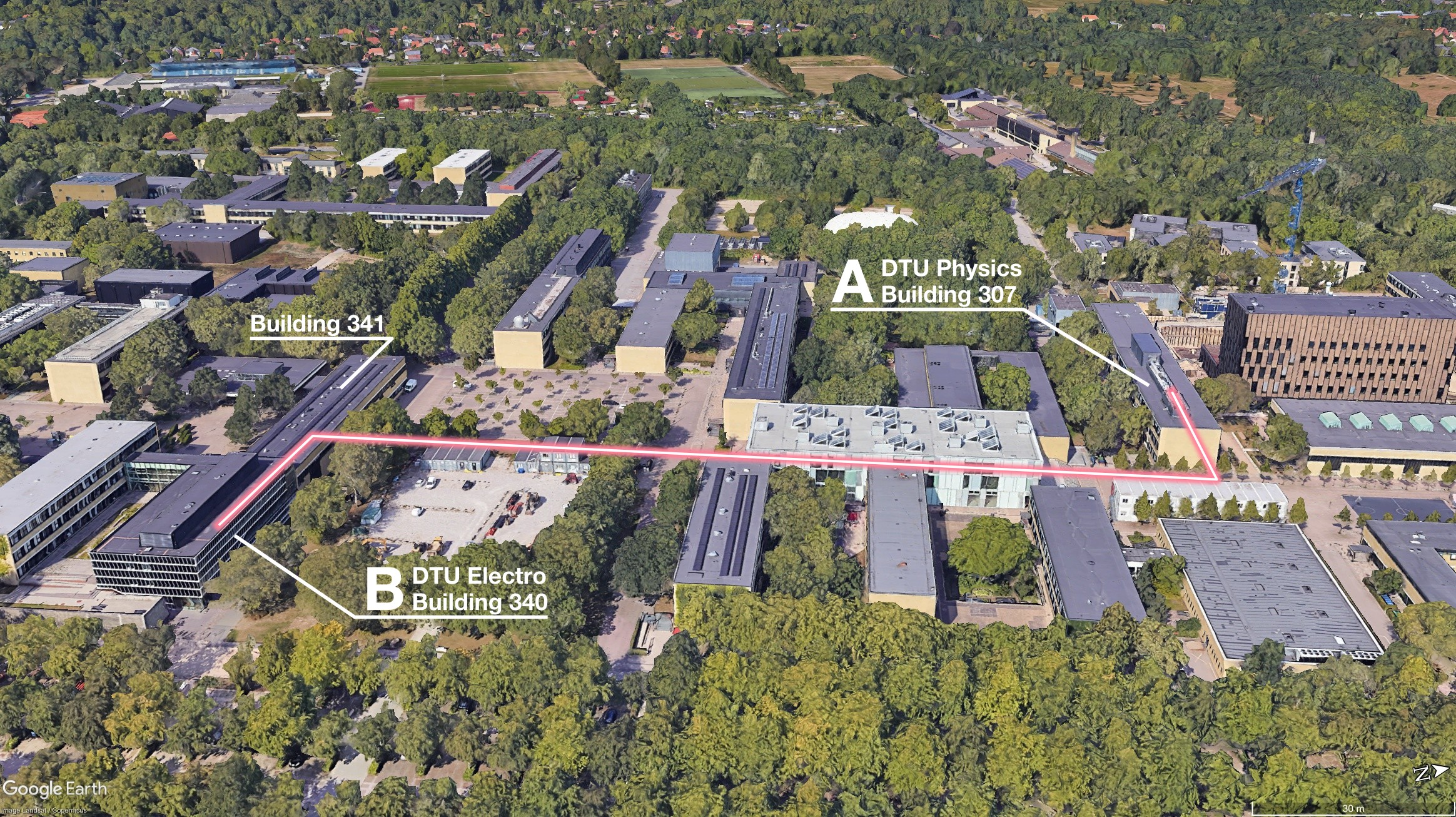}\\
    \vspace{1mm}
    \includegraphics[width=.75\linewidth]{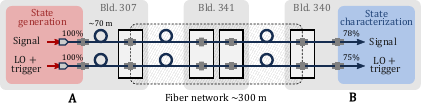}
    \caption{The Schr\"odinger kitten states are generated in a basement lab of DTU's building 307 (A) and distributed through $\sim$\SI{370}{m} of fiber installed in DTU's tunnel system -- as roughly indicated by the pink path -- to a ground-level technical room in building 340 (B) via a switchboard in building 341. 
    The optical link is made up of five SC-connectorized single-mode fiber patches. The central three patches of about \SI{300}{m} length (indicated by the hashed block in the diagram) are part of the already deployed campus network, while the first and last patches were set up for this demonstration. Due to the various connections at the intermediate stations, the optical states experience a loss of around 22\% during transmission.\\
    \textsl{Map data: Google, Landsat / Copernicus.}}
    \label{fig:campus}
\end{figure*}

Here, we take that crucial step by generating the non-Gaussian \textsl{Schr\"odinger kitten} states at the \SI{1550}{nm} telecom wavelength and transmitting them through a deployed fiber network between buildings separated by approximately \SI{250}{m} in straight-line distance and $\sim$\SI{370}{m} actual fiber distance. They traverse the ambient environment of the tunnel system beneath the Technical University of Denmark (DTU), as illustrated in Fig.~\ref{fig:campus}. 
Specifically, we generate non-Gaussian states based on photon-subtracted squeezed vacuum states in our optical laboratory and measure the received states in another building using a portable quantum state tomography station. While loss and phase noise are affecting the transmitted states, the quantum non-Gaussianity and even the Wigner negativity (when corrected for detection loss) of the states survive. 

The paper is organized as follows: Section II provides a detailed description of the experimental scheme used for the generation and measurement of non-Gaussian kitten states. Section III presents the transmission protocol and the main results of the paper. Section IV concludes with a summary of our findings and their potential impact on the future of quantum networking and computing.

\section{State generation and characterization} 

\subsection{Photon subtraction and kitten states}
The Schr\"odinger cat state is a superposition of two opposite-phase coherent states. A small optical Schr\"odinger cat state (typically called \textsl{kitten state}) can be generated by subtracting a single photon from a squeezed vacuum state \cite{Dakna1997} (hence the alternative, more formal name \textsl{photon-subtracted squeezed vacuum state}). 
Photon subtraction is usually implemented by splitting off a small part of the light and directing it to a photon counter, whose detection event heralds the removal of one photon from the main part of the beam. This process has been demonstrated multiple times \cite{Lvovsky2020-ProductionApplicationsNonGaussiana} and has, for example, been applied to the implementation of probabilistic quantum computing gates \cite{Tipsmark2011,Blandino2012a}, entanglement distillation \cite{takahashi_entanglement_2010}, fundamental tests of quantum mechanics \cite{parigi_probing_2007}, as well as generation of cat state entanglement \cite{Ourjoumtsev2009}, hybrid entanglement \cite{jeong_generation_2014,morin_remote_2014} and cat state qubits \cite{Neergaard-Nielsen2010}, while the resulting kitten state has been used for tele-amplification of coherent states \cite{neergaard-nielsen_quantum_2013} and itself been the object of quantum teleportation \cite{Lee2011g}, amplification by breeding \cite{Sychev2017,Konno2024}, and in-line squeezing \cite{Wang2022c,Yoshida2025}. 
The first realizations of kitten states were carried out with pulsed \SI{850}{nm} light \cite{Ourjoumtsev2006} and with continuous-wave \SI{852}{nm} light \cite{Neergaard-Nielsen2006} using a Titanium-Sapphire laser as a source for the experiment, and later it was also demonstrated at \SI{1064}{nm} using a Nd:YAG laser \cite{Grebien2022}. 
However, for quantum communication applications, it is important to be able to produce the state at a telecom wavelength such as \SI{1550}{nm} to ensure long-distance transmission in optical fibers with minimal loss. The efficient subtraction of a single photon from a squeezed state at \SI{1550}{nm} usually requires a cryogenic single-photon detector to lower the dark count rate, although it can be remedied by frequency conversion as demonstrated in \cite{Baune2017}. The direct subtraction of a single photon from squeezed vacuum at \SI{1550}{nm} was carried out first in \cite{Namekata2010}, with a Wigner-negativity in \cite{Breum2020}, and subsequently refined in \cite{Kawasaki2022} and \cite{Takase2022}.

\begin{figure*}
    \centering
    \includegraphics[width=.75\linewidth]{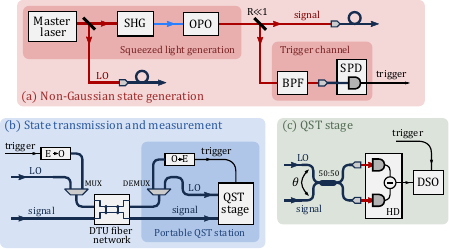}
    \caption{Experimental setup. SHG: second harmonic generation, OPO: optical parametric oscillator, BPF: bandpass filter, SPD: single photon detection, LO: local oscillator, E$\rightarrow$O: electrical to optical converter, O$\rightarrow$E: optical to electrical converter, (DE)MUX: optical (de)multiplexer, QST: quantum state tomography, HD: homodyne detection, DSO: digital storage oscilloscope. Please see main text for details.}
    \label{fig:setup}
\end{figure*}

\subsection{Generation of photon-subtracted squeezed states in the telecom band}
For generating \SI{1550}{nm} kitten states, we follow a similar approach to that of several of the earlier works mentioned above.
Our experimental setup is outlined in Fig. \ref{fig:setup}a and described in some detail in the following, while a more complete description is given in App.\ \ref{ch:app-exp}.
The master laser source is a low noise NKT Photonics fiber seed laser, whose output is amplified by an erbium-doped fiber amplifier and then frequency-doubled to \SI{775}{nm} in a separate second-harmonic generator (SHG). Squeezed states are produced in an optical parametric oscillator (OPO), which is identical to the SHG. It consists of a \SI{15}{mm} periodically poled KTP crystal in a bow-tie shaped ring cavity. The OPO has a free spectral range (FSR) of \SI{1.0}{GHz}, a half-width-half-max bandwidth of \SI{8.0}{MHz}, a pump threshold power of $\sim$\SI{800}{mW}, and an escape efficiency of 97\%. Details about the squeezed light setup can be found in \cite{Guo2020,Breum2020}. In this experiment, we inject a relatively weak pump of \SI{50}{mW}.

A photon is subtracted by reflecting a small fraction of the squeezed state on a highly asymmetric beam splitter (97/3) and detecting the presence of a single photon. 
Due to the broad phase-matching bandwidth of the nonlinear process, the output of the optical parametric amplifier is spectrally highly multimode, so a pair of optical filters is used to select only the central, degenerate mode prior to the photon detection. The first filter is a standard narrow-band free-space Fabry-Perot cavity with a bandwidth of \SI{78}{MHz} and an FSR of \SI{125}{GHz}. After coupling to single-mode fiber, the second filtering is done by a \SI{100}{GHz} DWDM (dense wavelength division multiplexing) filter. In combination, they efficiently select only the central mode at \SI{1550.12}{nm} of the frequency comb emitted from the squeezing cavity. 
The photon counter is a fiber-coupled superconducting nanowire single photon detector (SNSPD) based on a NbTiN-nanowire cavity stack operating at \SI{2}{K} \cite{Miki2013,Yamashita2013}. Upon the absorption of a photon, a voltage pulse is formed and subsequently amplified and sent through a discriminator to produce a clear signal to be used as trigger for the quantum state tomography. 
We measure a photon detection efficiency of around 60\% and typical dark count rates in the \qtyrange[range-units=single]{10}{50}{Hz} range. The total transmission through the two filters (and an acousto-optical modulator, see below) is around 30\%, resulting in a total efficiency of the trigger channel of about 20\%. 

We phase lock the optical cavities as well as the parametric gain of the squeezing operation and the measurement angle of the homodyne detector using standard feedback schemes implemented by RedPitaya STEMlab 125-14 FPGA modules using the PyRPL package \cite{Neuhaus2024}. The cavity locks require the use of bright, modulated auxiliary beams to create an error signal for feedback control. These beams will, however, saturate the photon counter (and also degrade the squeezing at low frequencies due to laser noise) and therefore cannot be active when the actual photon subtraction experiment is carried out. We therefore use a lock-measure method where all phase locks are active during a \textsl{lock} period, followed by a \textsl{measure} period where the locking beams are turned off using acousto-optical modulators, while the feedback signals are held constant and data are acquired within a period short enough that phases do not drift significantly. The lock period is set to \SI{7.0}{ms} while the measure period is \SI{2.5}{ms} with \SI{0.2}{ms} and \SI{0.3}{ms} of dead time before and after the measure period, respectively, for a repetition rate of \SI{100}{Hz}.    

\subsection{Quantum state tomography}

To characterize the conditionally generated quantum state prior to its transmission in the fiber network, we locally perform homodyne tomography immediately after coupling to a single-mode fiber with a coupling efficiency of 89\%. The homodyne detector (Fig.~\ref{fig:setup}c) consists of a fiber beam-splitter and a free-space balanced detector which we focus onto with GRIN lenses. 
The balanced detector has a \SI{40}{MHz} bandwidth and an average clearance within the spectral range of the squeezing of \SI{15}{dB} between shot noise and electronic noise levels using \SI{2}{mW} of optical local oscillator power. The estimated efficiency of the homodyne setup is $\sim$88\%. This includes propagation efficiency of 95\%, interference contrast of 99\%, diode efficiency of 97\%, and noise-equivalent efficiency \cite{appel_electronic_2007} of 97\%. 

\begin{figure}
    \centering
    \includegraphics[width=\linewidth]{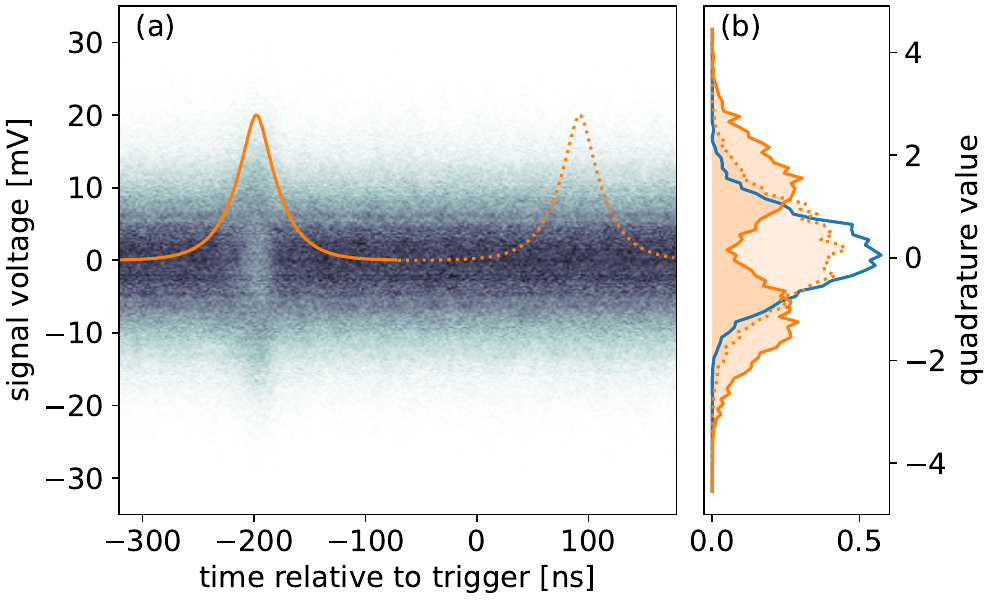}
    \caption{(a) Raw homodyne traces as a function of time relative to the trigger signal. The elevated noise region preceding the trigger corresponds to the temporal mode where a single photon has been subtracted from the squeezed vacuum state, leading to the generation of the non-Gaussian kitten state. (b) Histogram of quadrature measurements for the anti-squeezed p-quadrature using the optimal temporal mode function $f(t)$. The solid orange histogram corresponds to the photon-subtracted squeezed state, while the dotted histogram shows the reference distribution from the non-subtracted squeezed vacuum. Both distributions are normalized to the shot-noise level, which itself is indicated by the vacuum state histogram in solid blue.}
    \label{fig:quadrature_extraction}
\end{figure}

For the tomographic state analysis, we generate the same state 30,000 times and record on an oscilloscope its marginal quadrature distribution 5000 times at \SI{30}{\degree} intervals between \SI{0}{\degree} and \SI{150}{\degree} in phase space, as determined by the phase of the local oscillator.
The measurement window of a single trace, centered around a trigger click, was set to \SI{1}{\micro s} with a sampling rate of \SI{500}{MS/s}, yielding a time resolution of \SI{2}{ns}, which is sufficient for resolving the temporal correlations determined by the OPO bandwidth. 
Due to a longer optical and electronic path for the trigger (the SNSPD is located in a different room), the photon-subtracted state arrives a few hundred nanoseconds earlier than the trigger signal. This is visible in Fig.~\ref{fig:quadrature_extraction}a, which shows aggregate raw traces straight out of the oscilloscope, as an increased noise level with non-Gaussian distribution at this point in time.

From each trace, we extract a single quadrature value by integrating over the raw voltage trace multiplied by a temporal mode function $f(t)$ that can be arbitrarily chosen \cite{Neergaard-Nielsen2008}. 
The optimal mode function can in principle be obtained by performing an eigenfunction decomposition of the temporal autocovariance matrix of the homodyne traces~\cite{morin_remote_2014}, which yields the mode with maximum signal-to-noise ratio. However, we find that we get very close to the optimum (in terms of negativity of the reconstructed Wigner function) by choosing a physically informed mode function,
\begin{equation}
    f(t) = \frac{e^{-\gamma|t-t_0|}}{\gamma} - \frac{e^{-\kappa|t-t_0|}}{\kappa} \ ,
\end{equation}
where $t_0$ is the temporal offset, $\gamma/2\pi = 8.0\ \text{MHz}$ is the OPO bandwidth, and $\kappa/2\pi = 30\ \text{MHz}$ is an effective filtering bandwidth. This mode function represents a smoothening of the temporal correlations of the OPO output, which is a double-sided exponential. The smoothening is a combined effect of the optical trigger filtering, the limited bandwidth of the balanced detector, and a \SI{35}{MHz} low-pass filter that we inserted in the detector output.
A histogram of the extracted quadrature values for the anti-squeezed $p$ quadrature is shown in Fig.~\ref{fig:quadrature_extraction}b.
From the same data and using the same mode function, but shifted in time to be localized away from the photon-subtracted temporal region, we can extract quadrature values for the non-photon-subtracted squeezed vacuum state which is continuously produced from the OPO. This is illustrated by the dotted curves in Fig.~\ref{fig:quadrature_extraction}.
We normalize each of these curves to the shot noise level, obtained by recording the homodyne signal when blocking the input light and extracting the vacuum noise within the chosen temporal mode.
From the normalized, sampled quadrature distributions at six different phase space angles, we reconstruct the estimated density matrix and Wigner function of the generated state, using the standard Maximum Likelihood algorithm \cite{Lvovsky2009a}. In this reconstruction, we can choose to include a correction for the 12\% loss that can be attributed purely to the limited detection efficiency.

\begin{figure*}
    \centering
    \includegraphics[width=.32\linewidth]{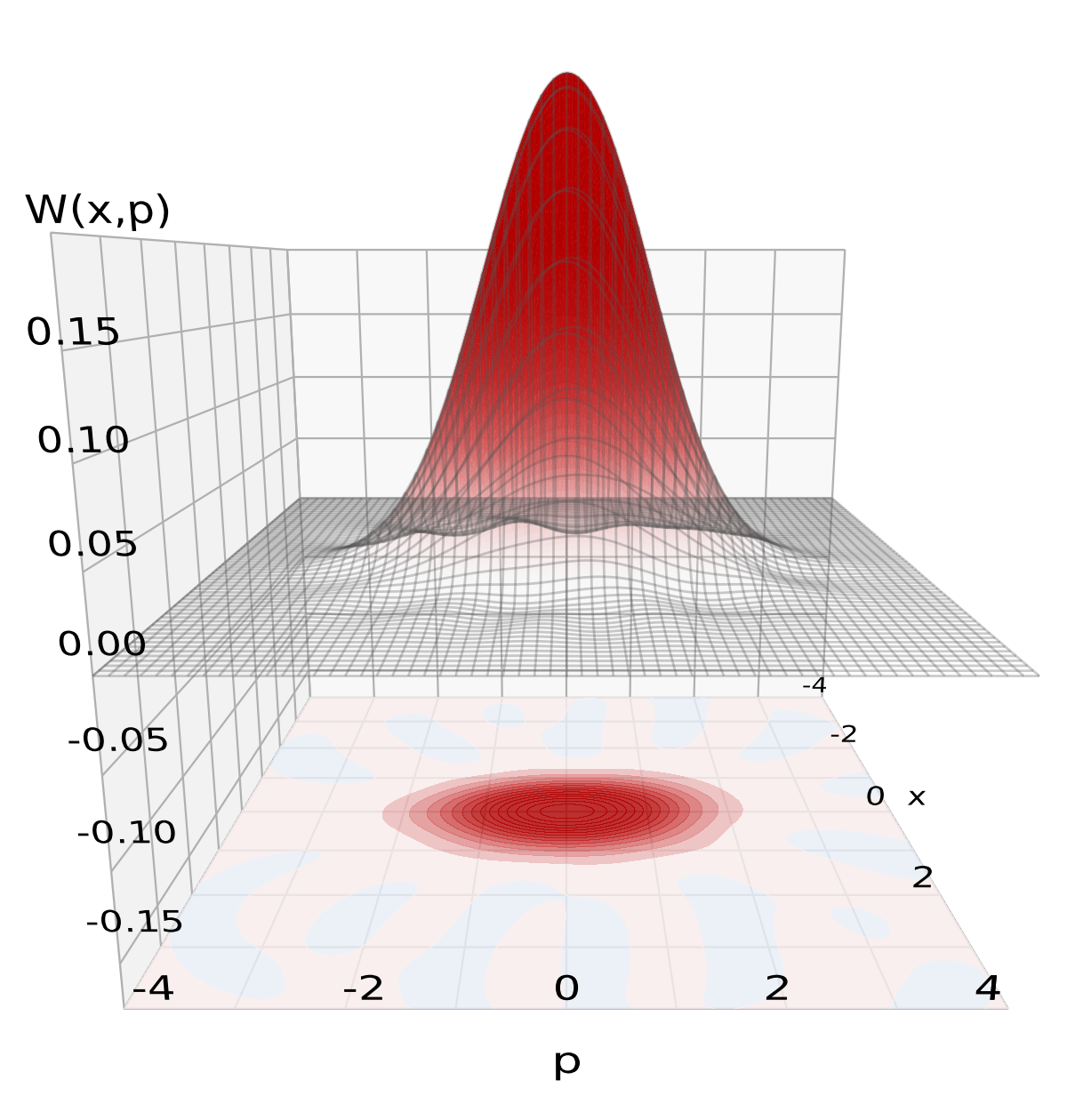}
    \includegraphics[width=.32\linewidth]{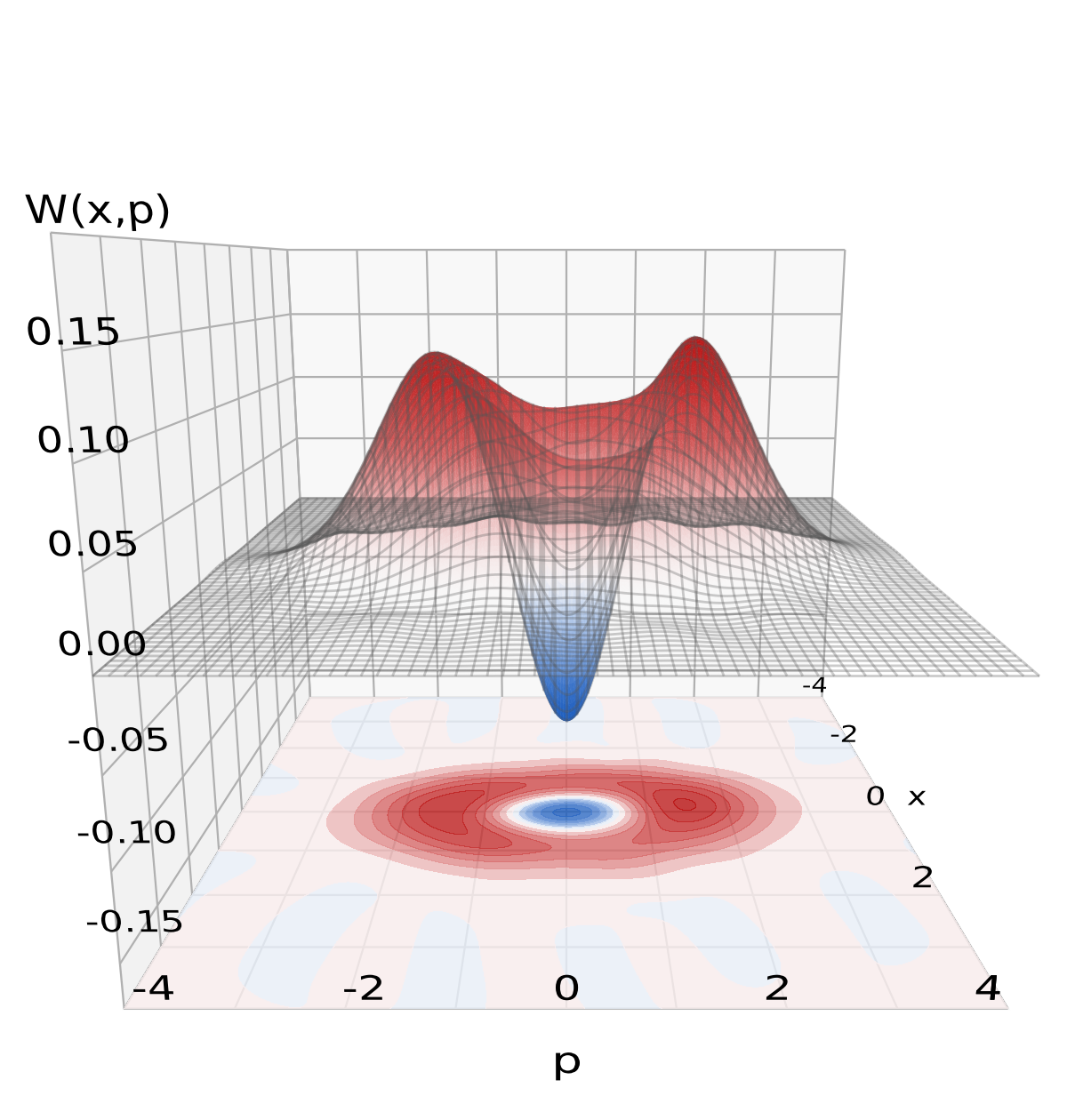}
    \includegraphics[width=.32\linewidth]{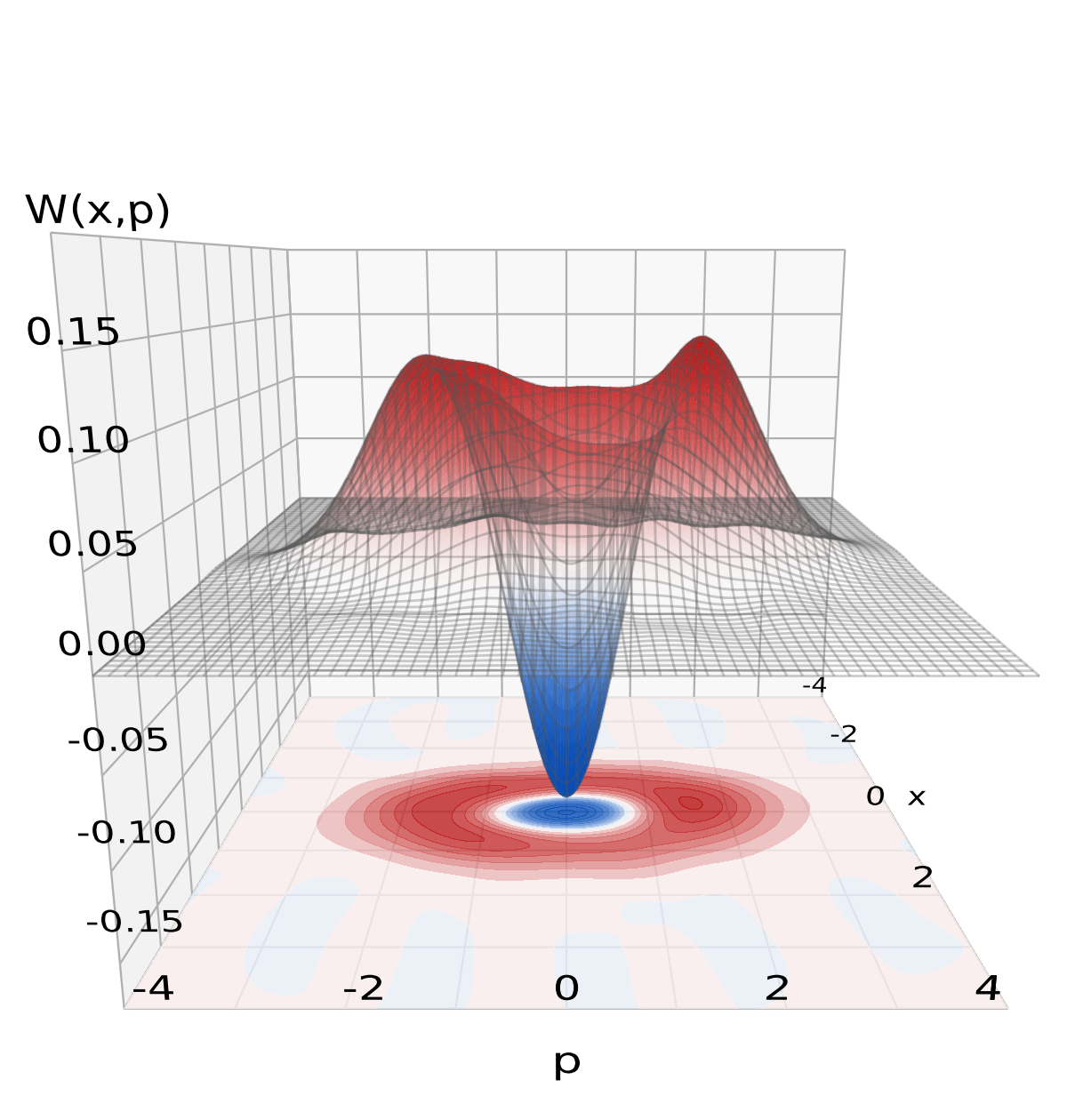}
    \caption{Reconstructed Wigner functions of the generated states. Wigner functions of the squeezed vacuum state (left), the photon-subtracted squeezed state (middle), and the loss-corrected photon-subtracted state (right) measured locally at the source in building A. The initial squeezed vacuum state shows Gaussian statistics, while the photon-subtracted state exhibits a pronounced dip in the Wigner function, $W(0,0) = -0.108(5)$, a hallmark of its non-Gaussian nature. Correcting for the 12\% detection loss further enhances the negativity to $W(0,0) = -0.164(4)$.}
    \label{fig:wigners307}
\end{figure*}

The reconstructed Wigner functions for the squeezed vacuum state, the photon-subtracted (kitten) state, and the loss-corrected kitten state are shown in Fig.~\ref{fig:wigners307}. The initial squeezed vacuum state exhibits a squeezing variance of -2.0 dB and anti-squeezing of 2.4 dB. After photon subtraction, we observe a clear dip in the Wigner function with a minimum value of $W(0,0)=-0.108(5)$, using the convention $\hbar=1$ (i.e.\ with the Wigner function confined within the interval $[-1/\pi;1/\pi]$). 
The uncertainty is obtained by parametric bootstrapping, i.e. by resampling 50 new datasets from the reconstructed state's marginal distributions.
When correcting for the $12\%$ detection loss, the negativity improves to $W(0,0)=-0.164(4)$, demonstrating strong nonclassicality. The generated kitten state has a fidelity of $66\%$ with an ideal odd Schr\"odinger cat state of amplitude $\alpha=0.91$. 

\section{Transmitted non-Gaussian state}

\subsection{State transmission and measurement station}

The signals, i.e., the non-Gaussian states generated in building A, are coupled into a \SI{70}{m} long single-mode fiber that, via a switchboard, connects to a deployed optical link of more than \SI{300}{m}. The states are transmitted through the $\sim$300~m long link with $\sim$78\% transmission to the receiving station in building B (see Fig.~\ref{fig:campus}). 
There, it is fully characterized by reconstructing its Wigner function using homodyne quantum state tomography. To this end, we transmit – along with the Schr\"odinger kitten states, but in a separate fiber – a bright optical signal at \SI{1550}{nm} for use as a coherent local oscillator in the homodyne detection.
The homodyne signal is recorded by an oscilloscope which is triggered by the SNSPD photon detection that heralds the generation of a kitten state. To transfer this heralding trigger signal from building A to building B, the SNSPD output is converted to a \SI{1310}{nm} optical signal with an electrical-optical converter and subsequently multiplexed with the \SI{1550}{nm} local oscillator into a single fiber.
At the receiving station, these two auxiliary signals are demultiplexed and the optical trigger signal is converted back into an electrical signal (see Fig.~\ref{fig:setup}b).
The electrical-optical-electrical conversion introduces a delay of \SI{11.2}{ns}, which can be compensated for, and a total jitter of less than \SI{24}{ps}, which is of the same order as the SNSPD jitter but significantly shorter than the duration of the generated non-Gaussian state ($\sim$\SI{60}{ns}) as determined by the OPO bandwidth.
Due to a lack of available channels, we do not transmit a signal carrying the timing of the \SI{100}{Hz} lock-measure period. Since this is so slow, it is possible to recreate the timing window locally in building B by a separate function generator generating a square wave with frequency and phase manually adjusted to match the arrival times of the trigger events (which, apart from dark counts, only appear in the measure period). Since the function generators in buildings A and B are not clock-synchronized, the measurement windows eventually drift apart, after $\sim$10 minutes. Future work should therefore include a synchronization of this signal to ensure long-term operation.

The quantum state tomography in building B is carried out by a mobile measurement station that is fiber-coupled and fully self-contained, including complete phase control and data acquisition (see Fig.~\ref{fig:setup}c). 
The balanced detector and the overall scheme are identical to those used for local characterization at the state generation in building A, including the overall detection efficiency of 88\%, but everything is placed on a rolling cart positioned next to the final switchboard in building B.

\subsection{Transmitted non-Gaussian state}

\begin{figure*}
    \centering
    \includegraphics[width=.32\linewidth]{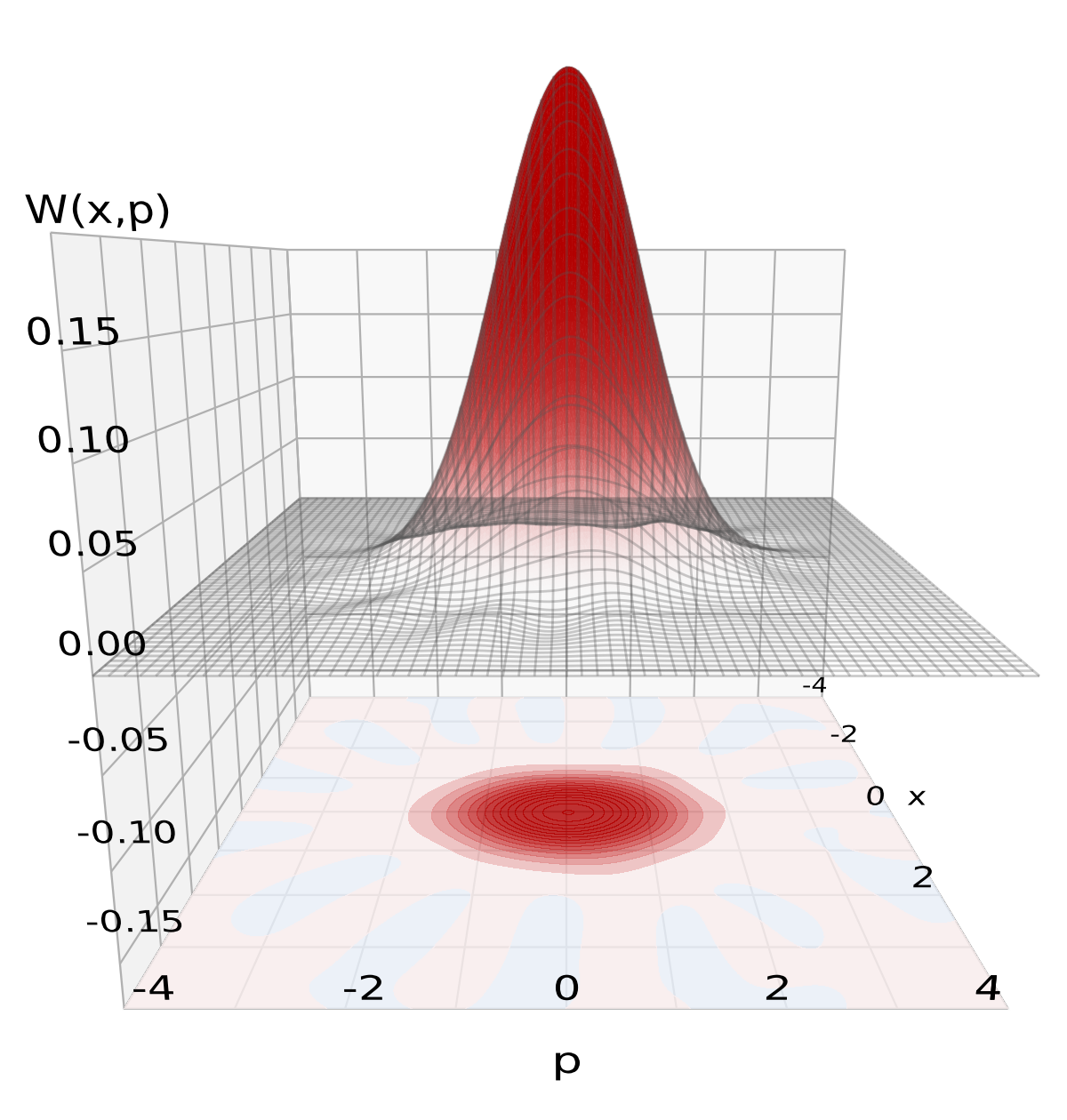}
    \includegraphics[width=.32\linewidth]{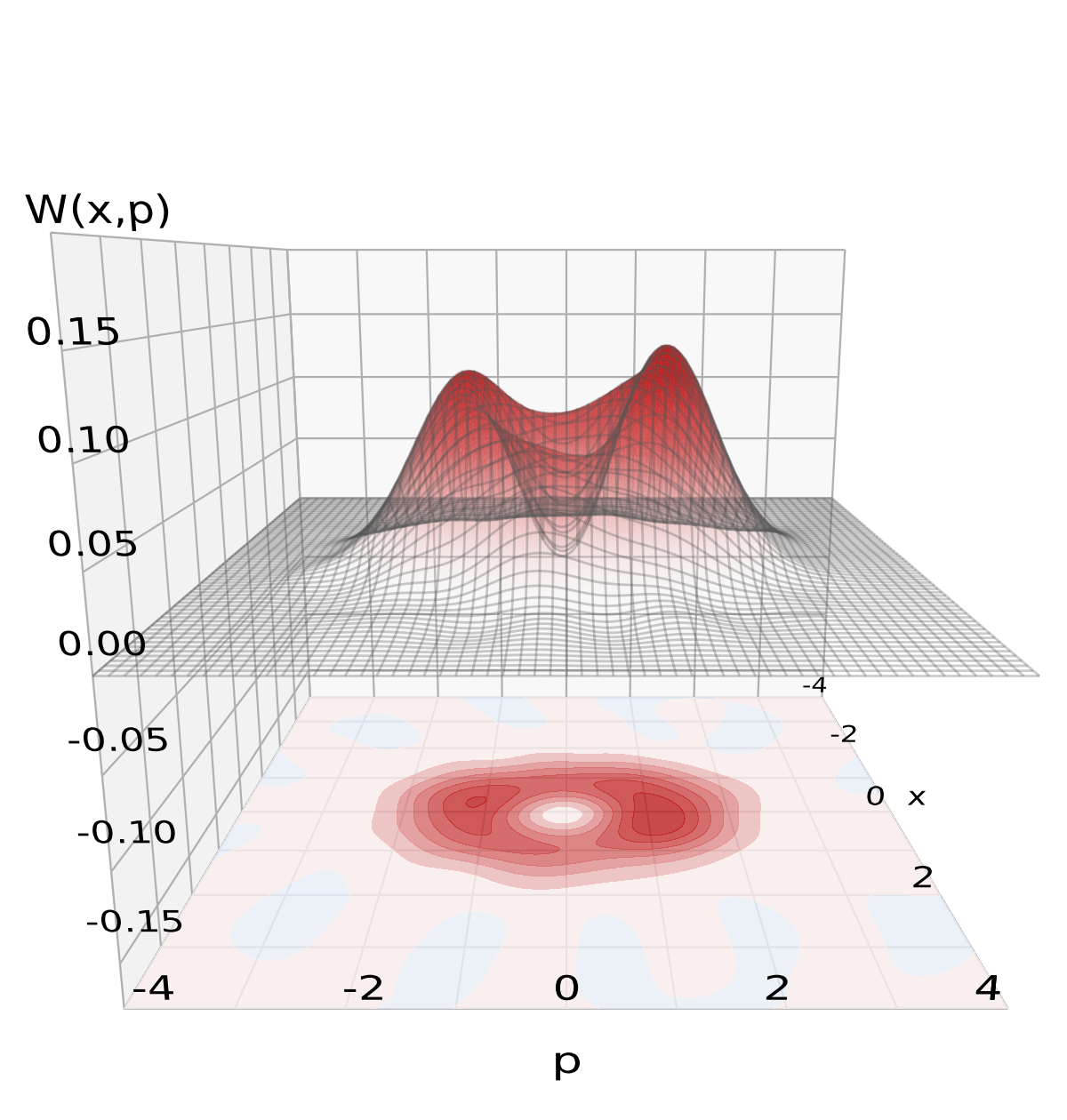}
    \includegraphics[width=.32\linewidth]{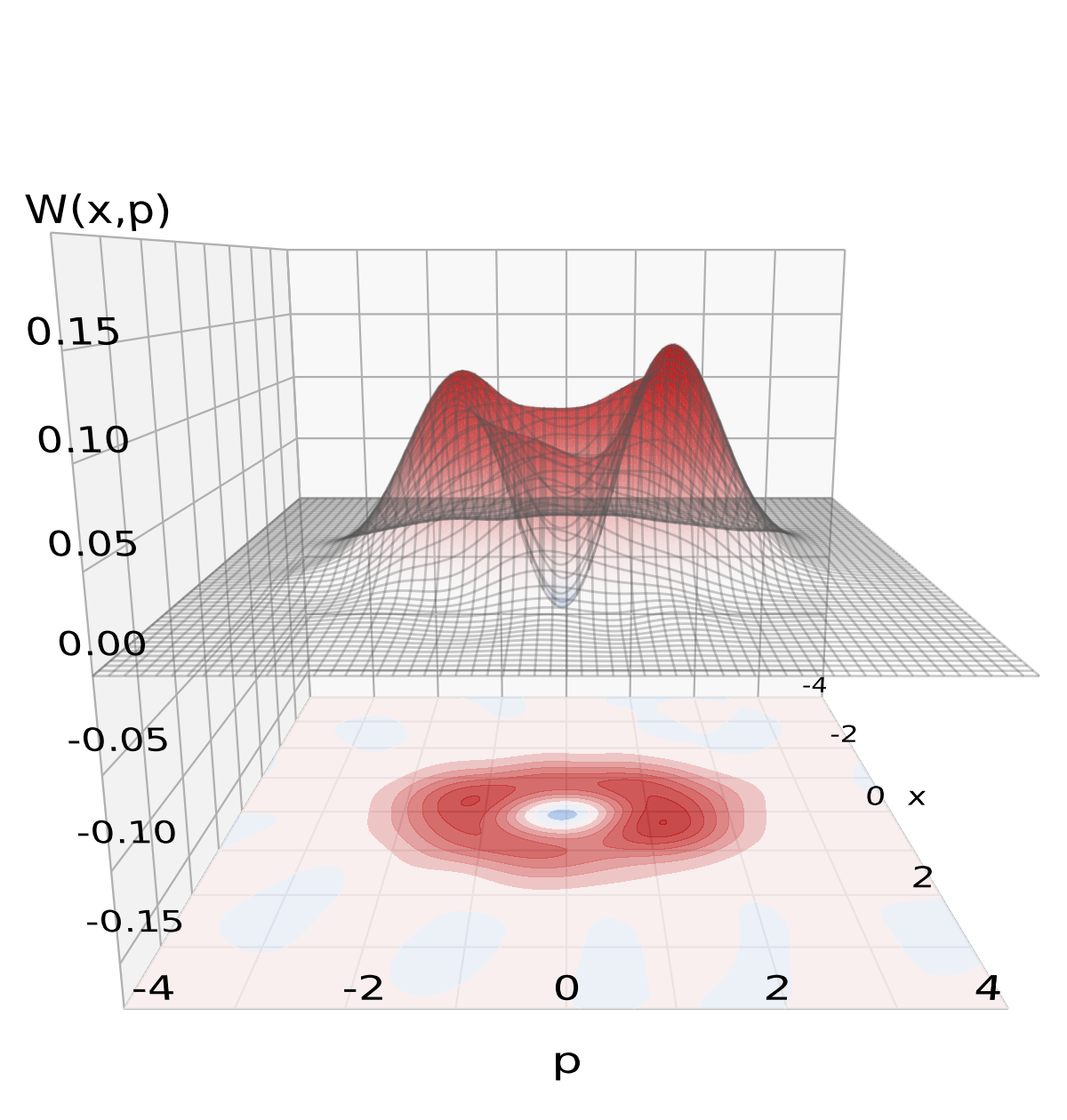}
    \caption{Reconstructed Wigner functions of the states transmitted across the campus fiber network. Wigner functions of the squeezed vacuum state (left), the photon-subtracted squeezed state (middle), and the loss-corrected photon-subtracted state (right) measured at the receiver in building B. When corrected for the 12\% inefficiency attributed purely to the homodyne detection at the receiving station, the transmitted state maintains a bit of negativity, with $W(0,0) = -0.028(4)$. Without this correction (middle), we have $W(0,0) = 0.006(5)$.}
    \label{fig:wigners340}
\end{figure*}

For reconstructing the Wigner functions of the transmitted states, we proceed exactly as for the reconstruction of the initial states in building A, with one exception:
To compensate for phase offsets in the receiver phase settings, we extract the actual local oscillator phases from the acquired data, as described in Appendix \ref{ch:app}, and use these as the quadrature angles.

After correction for 88\% detection efficiency and with the adjusted quadrature angles, we obtain the clearly non-Gaussian Wigner function illustrated in Fig.~\ref{fig:wigners340} (right). This state has a small, but unmistakable negativity at the center of $W(0,0)=-0.028(4)$, showing that we succeeded in transmitting a highly non-classical continuous-variable quantum state across a deployed fiber between the two buildings.
For reference, we also show the state without efficiency correction in Fig.~\ref{fig:wigners340} (middle), and the background squeezed vacuum state in the same figure on the left. Also, if we do not adjust the quadrature angles, but instead reconstruct using the intended 30$^\circ$ interval phases, we instead obtain a state with $W(0,0)=-0.020(5)$.
The locally measured and loss-corrected state was shown above to have a fidelity of 66\% with an ideal odd Schrödinger cat state of amplitude $\alpha=0.91$. After transmission through the deployed fiber link, this fidelity has decreased to 52\%, while the highest fidelity (53\%) is now obtained with a cat state of $\alpha=0.66$, consistent with the observed reduction in Wigner negativity due to losses and phase fluctuations during propagation.

\section{Conclusion and outlook}

In this study, we have achieved significant progress towards realizing the distribution of non-Gaussian continuous variable states within a functional real-world telecommunication setting. We established a reliable procedure for generating non-Gaussian states based on photon-subtracted squeezed vacuum states and for transmitting these states through an installed optical fiber network. This successful demonstration not only validates our methodology but also paves the way for future advancements and applications in this domain.

By enabling the distribution of non-Gaussian states in a practical telecommunication channel, our study provides a promising benchmark for developing advanced quantum networking protocols, such as quantum teleportation of non-Gaussian states, entanglement swapping and quantum repeaters based on bosonic codes. Importantly, the demonstrated survival of non-Gaussianity after transmission is a key prerequisite for realizing fundamental nonclassical effects such as Bell inequality violations and quantum steering over long distances. These capabilities significantly broaden the scope of high-dimensional, continuous-variable quantum information processing and open the door to future experiments exploring device-independent quantum communication. Looking ahead, one particularly promising direction for research is the integration of quantum error correction coding schemes for overcoming losses. These could potentially enable the faithful transmission of non-Gaussian states over long distances, addressing one of the major challenges in realizing a globally connected quantum internet.

\section{Acknowledgments}
This work was funded by the Danish National Research Foundation, Center for Macroscopic Quantum States (bigQ, DNRF0142), the Villum Foundation Young Investigator Programme (grant 10119), EU Horizon Europe (CLUSTEC, grant 101080173), and Innovation Fund Denmark/EU QuantERA (ClusSTAR, grant 3155-00024A).

\newpage 

\appendix

\section{Detailed experimental setup}
\label{ch:app-exp}

\begin{figure*}
    \centering
    \includegraphics[width=1\linewidth]{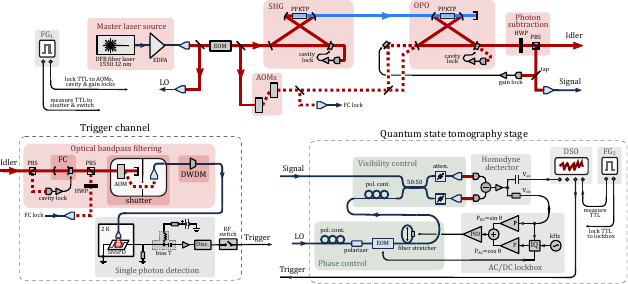}
    \caption{Detailed schematic of the experimental setup used for non-Gaussian state generation in Bld. 307 and for the quantum state tomography stage used in Bld. 341, as outlined in fig. \ref{fig:setup} in the main text. See App. \ref{ch:app-exp} for a detailed explanation of the setup. }
    \label{fig:app-schematic}
\end{figure*}

Fig.~\ref{fig:app-schematic} presents a schematic of the full experiment, each part of which is detailed below. Parts of this appendix overlap with the main text, but have been kept to make this description as complete as possible.

\textbf{State generation} - A master laser source consisting of a NKT Photonics Koheras BASIK X15 seed laser, which generates a low-noise single-mode \SI{1550.12}{nm} field, followed by a NKT Photonics erbium-doped fiber amplifier (EDFA), which amplifies the signal to $\sim\SI{2}{W}$, is used to run the full experiment. At first, a small portion of the laser field is tapped off and fiber-coupled to be sent to the quantum state tomography (QST) stage as the local oscillator (LO) for homodyne detection. 
Then, the main part of the \SI{1550.12}{nm} field is frequency doubled to \SI{775.06}{nm} in a second-harmonic generator (SHG) cavity. This field is used to pump an optical parametric oscillator (OPO) cavity, in order to produce squeezed light at \SI{1550.12}{nm} through degenerate spontaneous parametric down-conversion (SPDC). 
Both SHG and OPO cavities are identical and consist of a \SI{15}{mm} long periodically-poled (PP) KTP crystal in a bow-tie shaped ring cavity. They have a free spectral range (FSR) of \SI{1.0}{GHz}, a half-width-half-max bandwidth of \SI{8.0}{MHz}, a pump threshold power of $\sim$\SI{800}{mW}, and an escape efficiency of 97\%. 
After the OPO, a small portion of the squeezed light is tapped off as the idler/tapping field. The tapping ratio is controlled by a combination of half-wave plate (HWP) and polarization beamsplitter (PBS). 
Both SHG and OPO cavities are stabilized to the laser frequency using the Pound-Drever-Hall (PDH) locking technique using a \SI{28}{MHz} sideband tone generated by an EOM placed before the SHG cavity. For the OPO, a co-propagating probe beam is injected to act as a phase reference for the squeezed light and a counter-propagating locking beam is injected to obtain an error signal with sufficient signal-to-noise ratio. The error signal for the phase lock between the probe and squeezed light (called gain lock in Fig.~\ref{fig:app-schematic}), is obtained by detecting the parametric amplification of a \SI{50}{kHz} sideband on the probe beam through the OPO. The sideband is generated by the same piezo-mounted mirror used for feedback and the signal is detected by tapping off 1\% from the main signal arm after photon subtraction. Further details about this part of the setup can be found in \cite{Guo2020,Breum2020}.

\textbf{Trigger channel} - Due to the broad phase-matching bandwidth of the nonlinear process, the output of the OPO is highly spectrally multimode, so optical bandpass filtering (OBPF) is necessary to ensure that only the central, degenerate mode is coupled to the single photon detection. 
The OBPF consists of two filters and a shutter to protect the single photon detector. The first filter is a standard narrow-band free-space Fabry-Perot cavity with a bandwidth of \SI{78}{MHz} and an FSR of \SI{125}{GHz}. After coupling to single-mode fiber, the second filtering is done by a \SI{100}{GHz} dense wavelength division multiplexing (DWDM) filter. In combination, they efficiently select only the central mode at \SI{1550.12}{nm} of the frequency comb emitted from the OPO cavity. The photon counter is a fiber-coupled superconducting nanowire single photon detector (SNSPD) based on a NbTiN-nanowire cavity stack operating at \SI{2}{K} \cite{Miki2013,Yamashita2013}. Upon the absorption of a photon, a voltage pulse is formed and subsequently amplified and sent through a discriminator to produce a clear signal to be used as trigger for the quantum state tomography. We measure a photon detection efficiency of around 60\% and typical dark count rates in the \qtyrange[range-units=single]{10}{50}{Hz} range. The total transmission through the two filters and shutter is around 30\%, resulting in a total efficiency of the trigger channel of about 20\%. 
The filter cavity (FC) is stabilized with the PDH locking technique, using a counter-propagating locking beam in the orthogonal polarization. The choice of 3\% tapping ratio into the trigger channel is determined by the tradeoff between the added loss to the signal channel and the obtainable photon subtraction signal-to-noise ratio (total counts/dark counts) from the single photon detection. 

\textbf{Lock-measure scheme} - The cavity and phase locks require the use of bright, modulated auxiliary beams to create an error signal for feedback control. These beams will, however, saturate the photon counter (and also degrade the squeezing at low frequencies due to laser noise) and therefore cannot be active when the actual photon subtracting experiment is carried out. 
We therefore use a lock-measure scheme where all locks are active during a \textsl{lock} period, followed by a \textsl{measure} period where the locking and probe beams are turned off using acousto-optical modulators, while the feedback signals are held constant and data are acquired within a period short enough that phases do not drift significantly. The lock period is set to \SI{7.0}{ms} while the measure period is \SI{2.5}{ms} with \SI{0.2}{ms} and \SI{0.3}{ms} of dead time before and after the measure period, respectively, for a repetition rate of \SI{100}{Hz}. 
The corresponding optical chopping of the locking and probe beams is performed by a pair of acousto-optic modulators (AOM), which results in an amplitude modulation with \SI{60}{dB} extinction and \SI{0}{Hz} frequency offset. 
To further shield the photon counter, an additional AOM, which is only open during the measure period, is placed in the trigger channel before the fiber coupling inside an enclosure that minimizes stray light coupling to the fiber. 
Finally, an RF switch at the output of the single photon detection setup is used to gate the trigger signal, such that only single photon \textit{clicks} from the measure period are transmitted. All locking logic in the experiment is implemented by RedPitaya STEMlab 125-14 FPGA modules using the PyRPL package \cite{Neuhaus2024} and the TTL signals for the lock-measure scheme during state generation are produced by a standard function generator (FG).

\textbf{State transmission and measurement} - As depicted in Fig.~\ref{fig:setup} and described in the main text, the trigger signal is converted to a \SI{1310}{nm} optical pulse using a Highland Technologies electrical-to-fiberoptic (E2O) converter and combined with the \SI{1550.12}{nm} LO signal on a WDM multiplexer. 
The combined trigger and LO signal are then sent through the telecom network in a fiber channel adjacent to the fiber channel carrying the quantum signal. At the receiver side, in building 340, the trigger signal and LO signal are separated by a WDM demultiplexer, and the trigger signal is converted back to electrical pulses using a Highland Technologies fiberoptic-to-electrical (O2E) converter. These three signals are then coupled to the quantum state tomography (QST) stage, where the transmitted quantum state is measured by homodyne tomography. 

\begin{figure}
    \centering
    \includegraphics[width=0.9\linewidth]{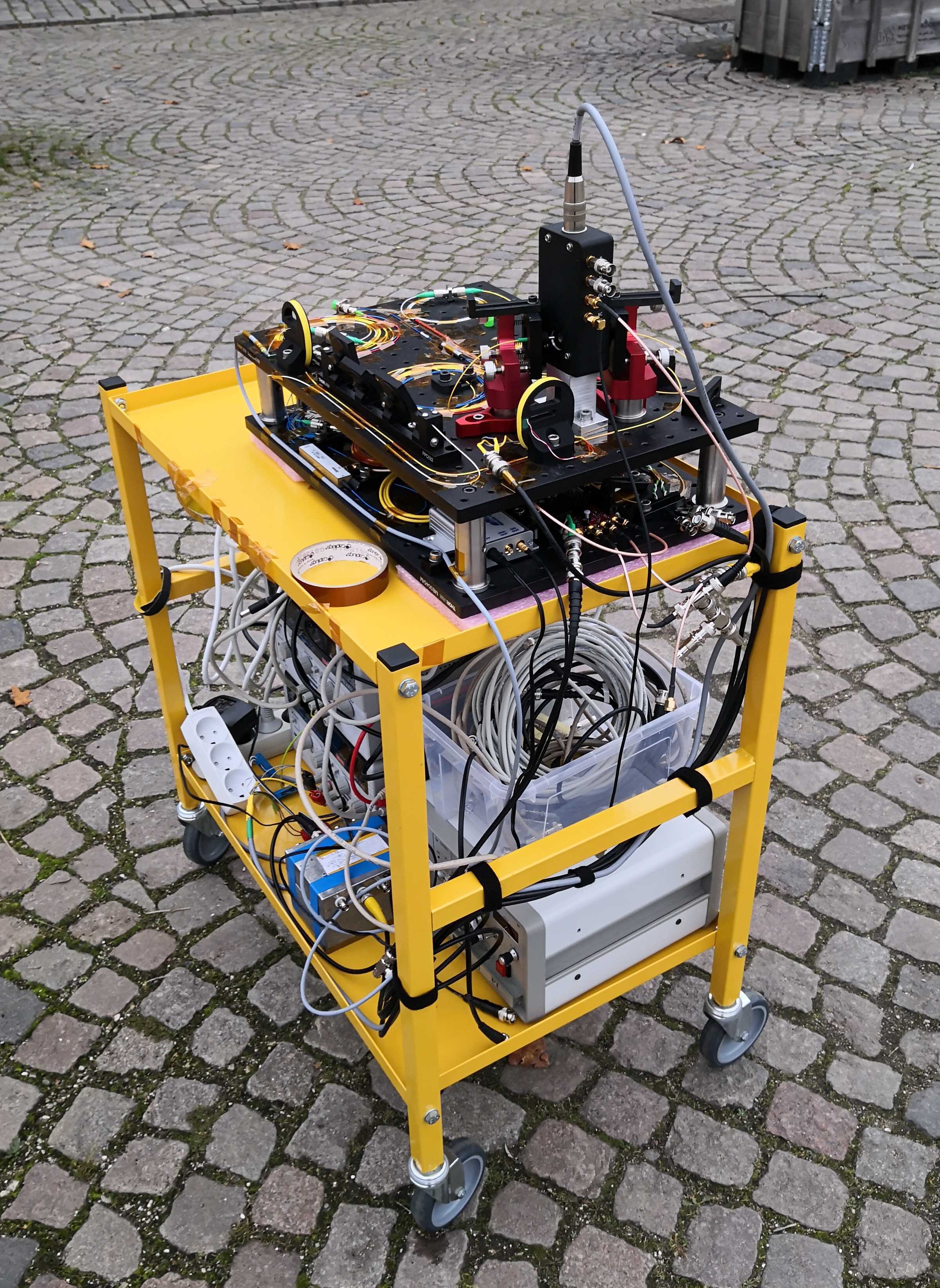}
    \caption{Picture of the fully self contained portable QST station on its way from building 307 to building 340 prior to the experimental runs. All optical components are mounted on the two-story 300x450 mm breadboard tower and the equipment below are lab power supplies, piezo drivers and function generator. Additional equipment not shown are digital-storage oscilloscope and PC.}
    \label{fig:app_pqst}
\end{figure}

\textbf{Quantum state tomography stage} - In order to characterize the signal state at the receiver side and verify its remaining non-Gaussianity, the Wigner function of the state is reconstructed using homodyne tomography. 
To perform this measurement locally in building 340, we construct a quantum state tomography (QST) stage consisting of a homodyne detector using high quantum efficiency photodiodes, a 12-bit real-time digital-storage oscilloscope (DSO) for recording the detector's AC-coupled output, a section to optimize the interference visibility between the LO and signal, and a section to lock the phase between the LO and signal to any arbitrary angle.
The visibility control section consists of an inline polarization controller to match the polarization state of the LO to the signal, and a 50:50 fiber coupler followed by inline attenuators in each output arm of the coupler to compensate for any slight asymmetry in coupling ratio or beam alignment onto the photodiodes. As explained in the main text, the total efficiency of the homodyne detection is estimated to 88\%. 
The phase control section consists of an inline polarization controller and polarizer to optimize the coupling from the SM fiber signal channel to a PM fiber coupled EOM, used to generate the phase reference sideband tone for locking, followed by a fiber stretcher used to effectuate the feedback phase control. The corresponding error signal is obtained using an AC/DC locking method described in the following. 

\textbf{Locking to arbitrary quadrature angles using AC/DC locking} - For the AC/DC locking scheme, both the AC and DC error signals are generated in parallel as depicted in Fig.~\ref{fig:app-schematic}. Here, the DC error signal is obtained directly as the DC output from the homodyne detector. The AC error signal is obtained by phase-modulating the LO at \SI{20}{kHz} with an EOM, followed by demodulating and low-pass filtering the same DC output of the homodyne detector. 
Both error signals are then first calibrated to have equal amplitude and signal-to-noise ratio and any offset removed, before being scaled according to $P_{DC}=\sin \theta$ and $P_{AC}=\cos \theta$, respectively, and finally summed. The resulting error signal $v_{\theta} = P_{DC}v_{DC} + P_{AC}v_{AC}$ is then sent to a PID controller. 
Thus, setting $\theta=0^o$ leads to pure AC locking and setting $\theta=90^o$ leads to pure DC locking, corresponding to measuring the squeezed and anti-squeezed quadrature, respectively. 
The full AC/DC lockbox is implemented using the PyRPL package \cite{Neuhaus2024} on a single RedPitaya STEMlab 125-14 FPGA module, allowing a full tomographic measurement series to be fully automated, including the error signal pre-calibration step and quadrature angle sweeping. 
As described in the following appendix section, this dependence of the quadrature angle on the precise and stable pre-calibration of the error signals unfortunately resulted in the locking angles differing from the preset values during measurement runs. 
Several reasons could cause this effect, including power fluctuations of the probe signal from gain lock instability or fiber coupling drift and power fluctuations of LO from polarization drift into the polarizer. 

As previously explained, a probe state is injected into the OPO and phase locked to the squeezed output in order to act as a phase reference for the non-Gaussian state at the receiver side, and it is also chopped according to the lock-measure scheme to protect the SNSPD. The AC/DC lockbox in building 340 therefore also needs to be provided with an identical lock-measure TTL signal in order to hold the quadrature feedback control constant at the correct times during the measure period. 
This is also crucial since the probe signal is gone during the measure period and the AC/DC lockbox would therefore otherwise try to provide feedback with no error signal and lose its lock. 
We generate the local lock TTL using a function generator identical to the one in building 307 and carefully detune its frequency to achieve a minimal phase drift of $\sim$ \SI{10}{\micro s} over 2 minutes between the original and local TTL signals. 

\section{Finding the correct quadrature angles}
\label{ch:app}

Due to relatively large phase fluctuations in the installed fiber and a phase locking system that was not perfectly adjusted, we ended up measuring the transmitted states at phase angles that were not exactly at the intended 30$^\circ$ intervals between 0$^\circ$ and 150$^\circ$. However, from the observed squeezing spectra, we were able to identify the actual quadrature angles, as described below. This helped us to more accurately reconstruct the transmitted states we presented in the main text.

Consider a Gaussian state of quadrature variance $V_x$ and $V_p$ of the orthogonal $x$- and $p$-quadratures, respectively. We define $\hat{x}$ as the squeezed quadrature of a squeezed thermal or vacuum state, such that $\langle \hat{x}\hat{p} +\hat{p}\hat{x} \rangle = 0$. The variance along the quadrature 
$\hat{q}_\theta=\hat{x}\cos\theta + \hat{p}\sin\theta$ is then
\begin{equation}
    V_\theta = V_x\cos^2\theta + V_p\sin^2\theta\;.
\end{equation}

When sampling $\hat{q}_\theta$ using homodyne detection, we expect $\theta$ to vary from sample to sample due to phase noise. Assuming the phase noise to be random with a Gaussian distribution of standard deviation $\sigma$, $G_\sigma(\phi)=\exp(-\phi^2/2\sigma^2)/\sqrt{2\pi\sigma^2}$, the resulting measured quadrature variance becomes
\begin{equation}\begin{aligned}\label{eq:app_Vm}
    V_\theta^{(\sigma)} &= \int_{-\infty}^\infty G_\sigma(\phi-\theta)V_\phi\,\text{d}\phi\\
    &=V_x\int_{-\infty}^\infty  G_\sigma(\phi-\theta)\cos^2\phi\,\text{d}\phi\\
    &\quad\quad+V_p\int_{-\infty}^\infty  G_\sigma(\phi-\theta)\sin^2\phi\,\text{d}\phi\\
    &=I\frac{V_x-V_p}{2}+\frac{V_x+V_p}{2}\;,
\end{aligned}\end{equation}
where
\begin{equation*}
    I=\int_{-\infty}^\infty G_\sigma(\phi-\theta)\cos2\phi\,\text{d}\phi= e^{-2\sigma^2}\cos2\theta\;,
\end{equation*}
and we used $\cos^2\phi=(1+\cos 2\phi)/2$ and $\sin^2\phi=(1-\cos2\phi)/2$. 
Inserting $I$ into eq.~\eqref{eq:app_Vm}, and using $\cos2\theta =\cos^2\theta-\sin^2\theta$, we get
\begin{equation}\begin{aligned}\label{eq:app_Vm2}
    V_\theta^{(\sigma)} &= e^{-2\sigma^2}\frac{V_x-V_p}{2}\cos2\theta+\frac{V_x + V_p}{2}\\
    &=e^{-2\sigma^2}\left(\frac{V_x+V_p}{2}+\frac{V_x-V_p}{2}\cos2\theta\right)\\
    &\quad\quad\quad\quad\quad\quad+\left(1-e^{-2\sigma^2}\right)\frac{V_x+V_p}{2}\\
    &=e^{-2\sigma^2}\left(V_x\cos^2\theta+ V_p\sin^2\theta\right)\\
    &\quad\quad+\left(1-e^{-2\sigma^2}\right)\frac{V_x+V_p}{2} \left(\cos^2\theta+\sin^2\theta\right)\\
    &=V_x^{(\sigma)}\cos^2\theta+V_p^{(\sigma)}\sin^2\theta\;,
\end{aligned}\end{equation}
with
\begin{equation*}
    V_{x(p)}^{(\sigma)} = \frac{1+e^{-2\sigma^2}}{2}V_{x(p)} + \frac{1-e^{-2\sigma^2}}{2}V_{p(x)}\;.
\end{equation*}
Notice that for no phase noise, $V_\theta^{(\sigma=0)}=V_\theta$, while for increasing phase noise $V_\theta^{(\sigma)}$ goes towards the average quadrature variance, $(V_x+V_p)/2$, as expected.

\begin{figure}
    \centering
    \includegraphics[width=0.99\linewidth]{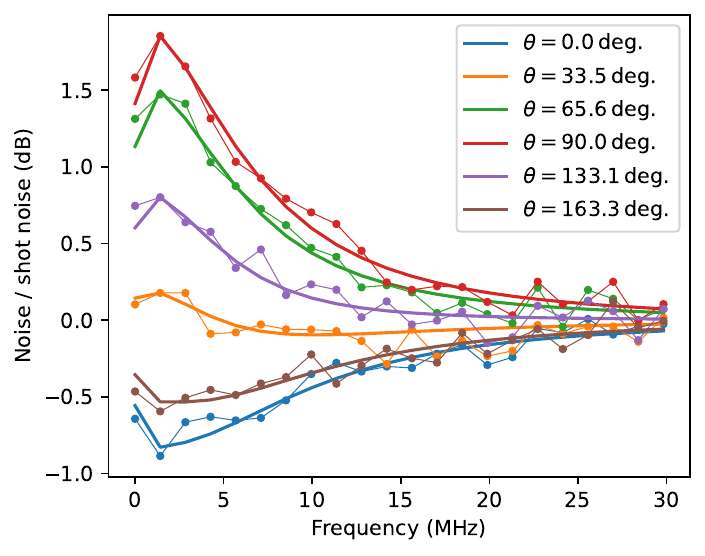}
    \caption{Squeezing spectra of time traces measured in Building B (dotted lines), and fitted squeezing model (solid lines). The squeezing spectra are extracted from the same time traces of which we perform state tomography, and the fitted quadrature angles are used state reconstruction.}
    \label{fig:app_spectrum}
\end{figure}
In Fig.~\ref{fig:app_spectrum} $V_\theta^{(\sigma)}$ is fitted to the measured squeezing spectra acquired in building 340 in order to determine the actual quadrature angles used in data acquisition for state tomography. The squeezing spectra are extracted from the same dataset used for state reconstruction, using the shifted temporal mode function as seen in Fig.~\ref{fig:quadrature_extraction}. Here we use
\begin{equation}\begin{aligned}
    V_x &= \frac{1}{2} - \frac{2\gamma\varepsilon\eta}{(\gamma+\varepsilon)^2 + (2\pi f)^2}\\
    V_p &= \frac{1}{2} + \frac{2\gamma\varepsilon\eta}{(\gamma-\varepsilon)^2 + (2\pi f)^2}\;,
\end{aligned}\end{equation}
where $\gamma=2\pi\times\SI{8.0}{MHz}$ is the OPO bandwidth, $\varepsilon$ is the OPO pump rate, and $\eta$ is the overall efficiency. $\varepsilon$ and $\eta$ are kept as free fitting parameters, together with the phase noise standard deviation, $\sigma$, and quadrature angle offsets at $\theta =30^\circ$, $60^\circ$, $120^\circ$ and $150^\circ$. 
The homodyne phase lock at $\theta=0^\circ$ and $90^\circ$ is implemented using AC and DC locks, respectively. Such AC and DC locks lock to orthogonal quadratures, and thus we include no phase offset in our fit at $\theta=0^\circ$ and $90^\circ$. 
At $\theta=30^\circ$, $60^\circ$, $120^\circ$, and $150^\circ$, the phase lock is implemented as a weighted AC/DC lock, where the weight of the AC and DC error signal is calibrated based on the desired quadrature angle and the error signal amplitudes. Since this calibration can be noisy and erroneous, phase offsets at these angles are included in the fitting. 

Finally, to take into account the frequency dependence of the homodyne shot noise clearance, a frequency dependent efficiency is included based on measured clearance. Note that the spectra at different quadrature angles are fitted all together in a single fit. The fitted (and fixed) parameters in Fig.~\ref{fig:app_spectrum} are:
\begin{equation*}\begin{aligned}
    \gamma &= 2\pi\times\SI{8}{MHz}\text{ (fixed)}\;,\\
    \varepsilon &= 2\pi\times\SI{1.74}{MHz}\;,\\
    \eta &= 0.462\;,\\
    \sigma &= 19.4^\circ\;,\\
    (\theta_0, \theta_{90}) &= (0^\circ, 90^\circ)\text{ (fixed)}\;,\\
    (\theta_{30},\theta_{60},\theta_{60},\theta_{60}) &= (33.5^\circ, 65.6^\circ, 133.1^\circ, 163.3^\circ)\;.\\
\end{aligned}\end{equation*}
The fitted quadrature angles, $\theta_{30}$, $\theta_{60}$, $\theta_{120}$, and $\theta_{150}$, are used for state reconstruction together with the fixed $\theta_0$ and $\theta_{90}$ in Fig.~\ref{fig:wigners340}.

\vspace{2cm}

\bibliography{remotecat}

\begin{thebibliography}{55}%
\makeatletter
\providecommand \@ifxundefined [1]{%
 \@ifx{#1\undefined}
}%
\providecommand \@ifnum [1]{%
 \ifnum #1\expandafter \@firstoftwo
 \else \expandafter \@secondoftwo
 \fi
}%
\providecommand \@ifx [1]{%
 \ifx #1\expandafter \@firstoftwo
 \else \expandafter \@secondoftwo
 \fi
}%
\providecommand \natexlab [1]{#1}%
\providecommand \enquote  [1]{``#1''}%
\providecommand \bibnamefont  [1]{#1}%
\providecommand \bibfnamefont [1]{#1}%
\providecommand \citenamefont [1]{#1}%
\providecommand \href@noop [0]{\@secondoftwo}%
\providecommand \href [0]{\begingroup \@sanitize@url \@href}%
\providecommand \@href[1]{\@@startlink{#1}\@@href}%
\providecommand \@@href[1]{\endgroup#1\@@endlink}%
\providecommand \@sanitize@url [0]{\catcode `\\12\catcode `\$12\catcode
  `\&12\catcode `\#12\catcode `\^12\catcode `\_12\catcode `\%12\relax}%
\providecommand \@@startlink[1]{}%
\providecommand \@@endlink[0]{}%
\providecommand \url  [0]{\begingroup\@sanitize@url \@url }%
\providecommand \@url [1]{\endgroup\@href {#1}{\urlprefix }}%
\providecommand \urlprefix  [0]{URL }%
\providecommand \Eprint [0]{\href }%
\providecommand \doibase [0]{https://doi.org/}%
\providecommand \selectlanguage [0]{\@gobble}%
\providecommand \bibinfo  [0]{\@secondoftwo}%
\providecommand \bibfield  [0]{\@secondoftwo}%
\providecommand \translation [1]{[#1]}%
\providecommand \BibitemOpen [0]{}%
\providecommand \bibitemStop [0]{}%
\providecommand \bibitemNoStop [0]{.\EOS\space}%
\providecommand \EOS [0]{\spacefactor3000\relax}%
\providecommand \BibitemShut  [1]{\csname bibitem#1\endcsname}%
\let\auto@bib@innerbib\@empty
\bibitem [{\citenamefont {Gottesman}\ \emph {et~al.}(2001)\citenamefont
  {Gottesman}, \citenamefont {Kitaev},\ and\ \citenamefont
  {Preskill}}]{Gottesman2001}%
  \BibitemOpen
  \bibfield  {author} {\bibinfo {author} {\bibfnamefont {D.}~\bibnamefont
  {Gottesman}}, \bibinfo {author} {\bibfnamefont {A.}~\bibnamefont {Kitaev}},\
  and\ \bibinfo {author} {\bibfnamefont {J.}~\bibnamefont {Preskill}},\
  }\bibfield  {title} {\bibinfo {title} {Encoding a qubit in an oscillator},\
  }\href {https://doi.org/10.1103/PhysRevA.64.012310} {\bibfield  {journal}
  {\bibinfo  {journal} {Phys. Rev. A}\ }\textbf {\bibinfo {volume} {64}},\
  \bibinfo {pages} {123101} (\bibinfo {year} {2001})}\BibitemShut {NoStop}%
\bibitem [{\citenamefont {Ralph}\ \emph {et~al.}(2003)\citenamefont {Ralph},
  \citenamefont {Gilchrist}, \citenamefont {Milburn}, \citenamefont {Munro},\
  and\ \citenamefont {Glancy}}]{Ralph2003}%
  \BibitemOpen
  \bibfield  {author} {\bibinfo {author} {\bibfnamefont {T.~C.}\ \bibnamefont
  {Ralph}}, \bibinfo {author} {\bibfnamefont {A.}~\bibnamefont {Gilchrist}},
  \bibinfo {author} {\bibfnamefont {G.~J.}\ \bibnamefont {Milburn}}, \bibinfo
  {author} {\bibfnamefont {W.~J.}\ \bibnamefont {Munro}},\ and\ \bibinfo
  {author} {\bibfnamefont {S.}~\bibnamefont {Glancy}},\ }\bibfield  {title}
  {\bibinfo {title} {Quantum computation with optical coherent states},\ }\href
  {https://doi.org/10.1103/PhysRevA.68.042319} {\bibfield  {journal} {\bibinfo
  {journal} {Phys. Rev. A}\ }\textbf {\bibinfo {volume} {68}},\ \bibinfo
  {pages} {042319} (\bibinfo {year} {2003})}\BibitemShut {NoStop}%
\bibitem [{\citenamefont {Albert}\ \emph {et~al.}(2018)\citenamefont {Albert},
  \citenamefont {Noh}, \citenamefont {Duivenvoorden}, \citenamefont {Young},
  \citenamefont {Brierley}, \citenamefont {Reinhold}, \citenamefont {Vuillot},
  \citenamefont {Li}, \citenamefont {Shen}, \citenamefont {Girvin},
  \citenamefont {Terhal},\ and\ \citenamefont {Jiang}}]{Albert2018}%
  \BibitemOpen
  \bibfield  {author} {\bibinfo {author} {\bibfnamefont {V.~V.}\ \bibnamefont
  {Albert}}, \bibinfo {author} {\bibfnamefont {K.}~\bibnamefont {Noh}},
  \bibinfo {author} {\bibfnamefont {K.}~\bibnamefont {Duivenvoorden}}, \bibinfo
  {author} {\bibfnamefont {D.~J.}\ \bibnamefont {Young}}, \bibinfo {author}
  {\bibfnamefont {R.~T.}\ \bibnamefont {Brierley}}, \bibinfo {author}
  {\bibfnamefont {P.}~\bibnamefont {Reinhold}}, \bibinfo {author}
  {\bibfnamefont {C.}~\bibnamefont {Vuillot}}, \bibinfo {author} {\bibfnamefont
  {L.}~\bibnamefont {Li}}, \bibinfo {author} {\bibfnamefont {C.}~\bibnamefont
  {Shen}}, \bibinfo {author} {\bibfnamefont {S.~M.}\ \bibnamefont {Girvin}},
  \bibinfo {author} {\bibfnamefont {B.~M.}\ \bibnamefont {Terhal}},\ and\
  \bibinfo {author} {\bibfnamefont {L.}~\bibnamefont {Jiang}},\ }\bibfield
  {title} {\bibinfo {title} {Performance and structure of single-mode bosonic
  codes},\ }\href {https://doi.org/10.1103/PhysRevA.97.032346} {\bibfield
  {journal} {\bibinfo  {journal} {Phys. Rev. A}\ }\textbf {\bibinfo {volume}
  {97}},\ \bibinfo {pages} {032346} (\bibinfo {year} {2018})}\BibitemShut
  {NoStop}%
\bibitem [{\citenamefont {Grimsmo}\ \emph {et~al.}(2020)\citenamefont
  {Grimsmo}, \citenamefont {Combes},\ and\ \citenamefont
  {Baragiola}}]{Grimsmo2020}%
  \BibitemOpen
  \bibfield  {author} {\bibinfo {author} {\bibfnamefont {A.~L.}\ \bibnamefont
  {Grimsmo}}, \bibinfo {author} {\bibfnamefont {J.}~\bibnamefont {Combes}},\
  and\ \bibinfo {author} {\bibfnamefont {B.~Q.}\ \bibnamefont {Baragiola}},\
  }\bibfield  {title} {\bibinfo {title} {Quantum computing with
  rotation-symmetric bosonic codes},\ }\href
  {https://doi.org/10.1103/PhysRevX.10.011058} {\bibfield  {journal} {\bibinfo
  {journal} {Phys. Rev. X}\ }\textbf {\bibinfo {volume} {10}},\ \bibinfo
  {pages} {011058} (\bibinfo {year} {2020})}\BibitemShut {NoStop}%
\bibitem [{\citenamefont {Menicucci}(2014)}]{Menicucci2014}%
  \BibitemOpen
  \bibfield  {author} {\bibinfo {author} {\bibfnamefont {N.~C.}\ \bibnamefont
  {Menicucci}},\ }\bibfield  {title} {\bibinfo {title} {Fault-tolerant
  measurement-based quantum computing with continuous-variable cluster
  states},\ }\href {https://doi.org/10.1103/physrevlett.112.120504} {\bibfield
  {journal} {\bibinfo  {journal} {Phys. Rev. Lett.}\ }\textbf {\bibinfo
  {volume} {112}},\ \bibinfo {pages} {120504} (\bibinfo {year}
  {2014})}\BibitemShut {NoStop}%
\bibitem [{\citenamefont {Larsen}\ \emph {et~al.}(2021)\citenamefont {Larsen},
  \citenamefont {Chamberland}, \citenamefont {Noh}, \citenamefont
  {{Neergaard-Nielsen}},\ and\ \citenamefont {Andersen}}]{Larsen2021}%
  \BibitemOpen
  \bibfield  {author} {\bibinfo {author} {\bibfnamefont {M.~V.}\ \bibnamefont
  {Larsen}}, \bibinfo {author} {\bibfnamefont {C.}~\bibnamefont {Chamberland}},
  \bibinfo {author} {\bibfnamefont {K.}~\bibnamefont {Noh}}, \bibinfo {author}
  {\bibfnamefont {J.~S.}\ \bibnamefont {{Neergaard-Nielsen}}},\ and\ \bibinfo
  {author} {\bibfnamefont {U.~L.}\ \bibnamefont {Andersen}},\ }\bibfield
  {title} {\bibinfo {title} {Fault-tolerant continuous-variable
  measurement-based quantum computation architecture},\ }\href
  {https://doi.org/10.1103/PRXQuantum.2.030325} {\bibfield  {journal} {\bibinfo
   {journal} {PRX Quantum}\ }\textbf {\bibinfo {volume} {2}},\ \bibinfo {pages}
  {030325} (\bibinfo {year} {2021})}\BibitemShut {NoStop}%
\bibitem [{\citenamefont {Tzitrin}\ \emph {et~al.}(2021)\citenamefont
  {Tzitrin}, \citenamefont {Matsuura}, \citenamefont {Alexander}, \citenamefont
  {Dauphinais}, \citenamefont {Bourassa}, \citenamefont {Sabapathy},
  \citenamefont {Menicucci},\ and\ \citenamefont {Dhand}}]{Tzitrin2021}%
  \BibitemOpen
  \bibfield  {author} {\bibinfo {author} {\bibfnamefont {I.}~\bibnamefont
  {Tzitrin}}, \bibinfo {author} {\bibfnamefont {T.}~\bibnamefont {Matsuura}},
  \bibinfo {author} {\bibfnamefont {R.~N.}\ \bibnamefont {Alexander}}, \bibinfo
  {author} {\bibfnamefont {G.}~\bibnamefont {Dauphinais}}, \bibinfo {author}
  {\bibfnamefont {J.~E.}\ \bibnamefont {Bourassa}}, \bibinfo {author}
  {\bibfnamefont {K.~K.}\ \bibnamefont {Sabapathy}}, \bibinfo {author}
  {\bibfnamefont {N.~C.}\ \bibnamefont {Menicucci}},\ and\ \bibinfo {author}
  {\bibfnamefont {I.}~\bibnamefont {Dhand}},\ }\bibfield  {title} {\bibinfo
  {title} {Fault-tolerant quantum computation with static linear optics},\
  }\href {https://doi.org/10.1103/PRXQuantum.2.040353} {\bibfield  {journal}
  {\bibinfo  {journal} {PRX Quantum}\ }\textbf {\bibinfo {volume} {2}},\
  \bibinfo {pages} {040353} (\bibinfo {year} {2021})}\BibitemShut {NoStop}%
\bibitem [{\citenamefont {Fukui}\ \emph {et~al.}(2021)\citenamefont {Fukui},
  \citenamefont {Alexander},\ and\ \citenamefont {{van Loock}}}]{Fukui2021}%
  \BibitemOpen
  \bibfield  {author} {\bibinfo {author} {\bibfnamefont {K.}~\bibnamefont
  {Fukui}}, \bibinfo {author} {\bibfnamefont {R.~N.}\ \bibnamefont
  {Alexander}},\ and\ \bibinfo {author} {\bibfnamefont {P.}~\bibnamefont {{van
  Loock}}},\ }\bibfield  {title} {\bibinfo {title} {All-{{Optical Long-Distance
  Quantum Communication}} with {{Gottesman-Kitaev-Preskill}} qubits},\ }\href
  {https://doi.org/10.1103/physrevresearch.3.033118} {\bibfield  {journal}
  {\bibinfo  {journal} {Phys. Rev. Research}\ }\textbf {\bibinfo {volume}
  {3}},\ \bibinfo {pages} {033118} (\bibinfo {year} {2021})}\BibitemShut
  {NoStop}%
\bibitem [{\citenamefont {Rozpedek}\ \emph {et~al.}(2021)\citenamefont
  {Rozpedek}, \citenamefont {Noh}, \citenamefont {Xu}, \citenamefont {Guha},\
  and\ \citenamefont {Jiang}}]{Rozpedek2021}%
  \BibitemOpen
  \bibfield  {author} {\bibinfo {author} {\bibfnamefont {F.}~\bibnamefont
  {Rozpedek}}, \bibinfo {author} {\bibfnamefont {K.}~\bibnamefont {Noh}},
  \bibinfo {author} {\bibfnamefont {Q.}~\bibnamefont {Xu}}, \bibinfo {author}
  {\bibfnamefont {S.}~\bibnamefont {Guha}},\ and\ \bibinfo {author}
  {\bibfnamefont {L.}~\bibnamefont {Jiang}},\ }\bibfield  {title} {\bibinfo
  {title} {Quantum repeaters based on concatenated bosonic and
  discrete-variable quantum codes},\ }\href
  {https://doi.org/10.1038/s41534-021-00438-7} {\bibfield  {journal} {\bibinfo
  {journal} {npj Quantum Inf.}\ }\textbf {\bibinfo {volume} {7}},\ \bibinfo
  {pages} {102} (\bibinfo {year} {2021})}\BibitemShut {NoStop}%
\bibitem [{\citenamefont {Wu}\ \emph {et~al.}(2022)\citenamefont {Wu},
  \citenamefont {Zhang},\ and\ \citenamefont {Zhuang}}]{Wu2022}%
  \BibitemOpen
  \bibfield  {author} {\bibinfo {author} {\bibfnamefont {B.~H.}\ \bibnamefont
  {Wu}}, \bibinfo {author} {\bibfnamefont {Z.}~\bibnamefont {Zhang}},\ and\
  \bibinfo {author} {\bibfnamefont {Q.}~\bibnamefont {Zhuang}},\ }\bibfield
  {title} {\bibinfo {title} {Continuous-variable quantum repeaters based on
  bosonic error-correction and teleportation: {{Architecture}} and
  applications},\ }\href {https://doi.org/10.1088/2058-9565/ac4f6b} {\bibfield
  {journal} {\bibinfo  {journal} {Quantum Science and Technology}\ }\textbf
  {\bibinfo {volume} {7}},\ \bibinfo {pages} {025018} (\bibinfo {year}
  {2022})}\BibitemShut {NoStop}%
\bibitem [{\citenamefont {Ourjoumtsev}\ \emph {et~al.}(2006)\citenamefont
  {Ourjoumtsev}, \citenamefont {{Tualle-Brouri}}, \citenamefont {Laurat},\ and\
  \citenamefont {Grangier}}]{Ourjoumtsev2006}%
  \BibitemOpen
  \bibfield  {author} {\bibinfo {author} {\bibfnamefont {A.}~\bibnamefont
  {Ourjoumtsev}}, \bibinfo {author} {\bibfnamefont {R.}~\bibnamefont
  {{Tualle-Brouri}}}, \bibinfo {author} {\bibfnamefont {J.}~\bibnamefont
  {Laurat}},\ and\ \bibinfo {author} {\bibfnamefont {P.}~\bibnamefont
  {Grangier}},\ }\bibfield  {title} {\bibinfo {title} {Generating {{Optical
  Schr{\"o}dinger Kittens}} for {{Quantum Information Processing}}},\ }\href
  {https://doi.org/10.1126/science.1122858} {\bibfield  {journal} {\bibinfo
  {journal} {Science}\ }\textbf {\bibinfo {volume} {312}},\ \bibinfo {pages}
  {83} (\bibinfo {year} {2006})}\BibitemShut {NoStop}%
\bibitem [{\citenamefont {{Neergaard-Nielsen}}\ \emph
  {et~al.}(2006)\citenamefont {{Neergaard-Nielsen}}, \citenamefont {Nielsen},
  \citenamefont {Hettich}, \citenamefont {M{\o}lmer},\ and\ \citenamefont
  {Polzik}}]{Neergaard-Nielsen2006}%
  \BibitemOpen
  \bibfield  {author} {\bibinfo {author} {\bibfnamefont {J.~S.}\ \bibnamefont
  {{Neergaard-Nielsen}}}, \bibinfo {author} {\bibfnamefont {B.~M.}\
  \bibnamefont {Nielsen}}, \bibinfo {author} {\bibfnamefont {C.}~\bibnamefont
  {Hettich}}, \bibinfo {author} {\bibfnamefont {K.}~\bibnamefont {M{\o}lmer}},\
  and\ \bibinfo {author} {\bibfnamefont {E.~S.}\ \bibnamefont {Polzik}},\
  }\bibfield  {title} {\bibinfo {title} {Generation of a superposition of odd
  photon number states for quantum information networks},\ }\href
  {https://doi.org/10.1103/PhysRevLett.97.083604} {\bibfield  {journal}
  {\bibinfo  {journal} {Phys. Rev. Lett.}\ }\textbf {\bibinfo {volume} {97}},\
  \bibinfo {pages} {083604} (\bibinfo {year} {2006})}\BibitemShut {NoStop}%
\bibitem [{\citenamefont {Larsen}\ \emph {et~al.}(2019)\citenamefont {Larsen},
  \citenamefont {Guo}, \citenamefont {Breum}, \citenamefont
  {{Neergaard-Nielsen}},\ and\ \citenamefont {Andersen}}]{Larsen2019}%
  \BibitemOpen
  \bibfield  {author} {\bibinfo {author} {\bibfnamefont {M.~V.}\ \bibnamefont
  {Larsen}}, \bibinfo {author} {\bibfnamefont {X.}~\bibnamefont {Guo}},
  \bibinfo {author} {\bibfnamefont {C.~R.}\ \bibnamefont {Breum}}, \bibinfo
  {author} {\bibfnamefont {J.~S.}\ \bibnamefont {{Neergaard-Nielsen}}},\ and\
  \bibinfo {author} {\bibfnamefont {U.~L.}\ \bibnamefont {Andersen}},\
  }\bibfield  {title} {\bibinfo {title} {Deterministic generation of a
  two-dimensional cluster state},\ }\href
  {https://doi.org/10.1126/science.aay4354} {\bibfield  {journal} {\bibinfo
  {journal} {Science}\ }\textbf {\bibinfo {volume} {366}},\ \bibinfo {pages}
  {369} (\bibinfo {year} {2019})}\BibitemShut {NoStop}%
\bibitem [{\citenamefont {Asavanant}\ \emph {et~al.}(2019)\citenamefont
  {Asavanant}, \citenamefont {Shiozawa}, \citenamefont {Yokoyama},
  \citenamefont {Charoensombutamon}, \citenamefont {Emura}, \citenamefont
  {Alexander}, \citenamefont {Takeda}, \citenamefont {Yoshikawa}, \citenamefont
  {Menicucci}, \citenamefont {Yonezawa},\ and\ \citenamefont
  {Furusawa}}]{Asavanant2019}%
  \BibitemOpen
  \bibfield  {author} {\bibinfo {author} {\bibfnamefont {W.}~\bibnamefont
  {Asavanant}}, \bibinfo {author} {\bibfnamefont {Y.}~\bibnamefont {Shiozawa}},
  \bibinfo {author} {\bibfnamefont {S.}~\bibnamefont {Yokoyama}}, \bibinfo
  {author} {\bibfnamefont {B.}~\bibnamefont {Charoensombutamon}}, \bibinfo
  {author} {\bibfnamefont {H.}~\bibnamefont {Emura}}, \bibinfo {author}
  {\bibfnamefont {R.~N.}\ \bibnamefont {Alexander}}, \bibinfo {author}
  {\bibfnamefont {S.}~\bibnamefont {Takeda}}, \bibinfo {author} {\bibfnamefont
  {J.-I.}\ \bibnamefont {Yoshikawa}}, \bibinfo {author} {\bibfnamefont {N.~C.}\
  \bibnamefont {Menicucci}}, \bibinfo {author} {\bibfnamefont {H.}~\bibnamefont
  {Yonezawa}},\ and\ \bibinfo {author} {\bibfnamefont {A.}~\bibnamefont
  {Furusawa}},\ }\bibfield  {title} {\bibinfo {title} {Time-domain multiplexed
  2-dimensional cluster state: Universal quantum computing platform},\ }\href
  {https://doi.org/10.1126/science.aay2645} {\bibfield  {journal} {\bibinfo
  {journal} {Science}\ }\textbf {\bibinfo {volume} {366}},\ \bibinfo {pages}
  {373} (\bibinfo {year} {2019})}\BibitemShut {NoStop}%
\bibitem [{\citenamefont {Fl{\"u}hmann}\ \emph {et~al.}(2019)\citenamefont
  {Fl{\"u}hmann}, \citenamefont {Nguyen}, \citenamefont {Marinelli},
  \citenamefont {Negnevitsky}, \citenamefont {Mehta},\ and\ \citenamefont
  {Home}}]{Fluhmann2019}%
  \BibitemOpen
  \bibfield  {author} {\bibinfo {author} {\bibfnamefont {C.}~\bibnamefont
  {Fl{\"u}hmann}}, \bibinfo {author} {\bibfnamefont {T.~L.}\ \bibnamefont
  {Nguyen}}, \bibinfo {author} {\bibfnamefont {M.}~\bibnamefont {Marinelli}},
  \bibinfo {author} {\bibfnamefont {V.}~\bibnamefont {Negnevitsky}}, \bibinfo
  {author} {\bibfnamefont {K.}~\bibnamefont {Mehta}},\ and\ \bibinfo {author}
  {\bibfnamefont {J.~P.}\ \bibnamefont {Home}},\ }\bibfield  {title} {\bibinfo
  {title} {Encoding a qubit in a trapped-ion mechanical oscillator},\ }\href
  {https://doi.org/10.1038/s41586-019-0960-6} {\bibfield  {journal} {\bibinfo
  {journal} {Nature}\ }\textbf {\bibinfo {volume} {566}},\ \bibinfo {pages}
  {513} (\bibinfo {year} {2019})}\BibitemShut {NoStop}%
\bibitem [{\citenamefont {{Campagne-Ibarcq}}\ \emph {et~al.}(2020)\citenamefont
  {{Campagne-Ibarcq}}, \citenamefont {Eickbusch}, \citenamefont {Touzard},
  \citenamefont {{Zalys-Geller}}, \citenamefont {Frattini}, \citenamefont
  {Sivak}, \citenamefont {Reinhold}, \citenamefont {Puri}, \citenamefont
  {Shankar}, \citenamefont {Schoelkopf}, \citenamefont {Frunzio}, \citenamefont
  {Mirrahimi},\ and\ \citenamefont {Devoret}}]{Campagne-Ibarcq2020}%
  \BibitemOpen
  \bibfield  {author} {\bibinfo {author} {\bibfnamefont {P.}~\bibnamefont
  {{Campagne-Ibarcq}}}, \bibinfo {author} {\bibfnamefont {A.}~\bibnamefont
  {Eickbusch}}, \bibinfo {author} {\bibfnamefont {S.}~\bibnamefont {Touzard}},
  \bibinfo {author} {\bibfnamefont {E.}~\bibnamefont {{Zalys-Geller}}},
  \bibinfo {author} {\bibfnamefont {N.~E.}\ \bibnamefont {Frattini}}, \bibinfo
  {author} {\bibfnamefont {V.~V.}\ \bibnamefont {Sivak}}, \bibinfo {author}
  {\bibfnamefont {P.}~\bibnamefont {Reinhold}}, \bibinfo {author}
  {\bibfnamefont {S.}~\bibnamefont {Puri}}, \bibinfo {author} {\bibfnamefont
  {S.}~\bibnamefont {Shankar}}, \bibinfo {author} {\bibfnamefont {R.~J.}\
  \bibnamefont {Schoelkopf}}, \bibinfo {author} {\bibfnamefont
  {L.}~\bibnamefont {Frunzio}}, \bibinfo {author} {\bibfnamefont
  {M.}~\bibnamefont {Mirrahimi}},\ and\ \bibinfo {author} {\bibfnamefont
  {M.~H.}\ \bibnamefont {Devoret}},\ }\bibfield  {title} {\bibinfo {title}
  {Quantum error correction of a qubit encoded in grid states of an
  oscillator},\ }\href {https://doi.org/10.1038/s41586-020-2603-3} {\bibfield
  {journal} {\bibinfo  {journal} {Nature}\ }\textbf {\bibinfo {volume} {584}},\
  \bibinfo {pages} {368} (\bibinfo {year} {2020})}\BibitemShut {NoStop}%
\bibitem [{\citenamefont {Larsen}\ \emph {et~al.}(2025)\citenamefont {Larsen},
  \citenamefont {Bourassa}, \citenamefont {Kocsis}, \citenamefont {Tasker},
  \citenamefont {Chadwick}, \citenamefont {{Gonz{\'a}lez-Arciniegas}},
  \citenamefont {Hastrup}, \citenamefont {{Lopetegui-Gonz{\'a}lez}},
  \citenamefont {Miatto}, \citenamefont {Motamedi}, \citenamefont {Noro},
  \citenamefont {Roeland}, \citenamefont {Baby}, \citenamefont {Chen},
  \citenamefont {Contu}, \citenamefont {Di~Luch}, \citenamefont {Drago},
  \citenamefont {Giesbrecht}, \citenamefont {Grainge}, \citenamefont
  {Krasnokutska}, \citenamefont {Menotti}, \citenamefont {Morrison},
  \citenamefont {Puviraj}, \citenamefont {Rezaei~Shad}, \citenamefont
  {Hussain}, \citenamefont {McMahon}, \citenamefont {Ortmann}, \citenamefont
  {Collins}, \citenamefont {Ma}, \citenamefont {Phillips}, \citenamefont
  {Seymour}, \citenamefont {Tang}, \citenamefont {Yang}, \citenamefont
  {Vernon}, \citenamefont {Alexander},\ and\ \citenamefont
  {Mahler}}]{larsen_integrated_2025}%
  \BibitemOpen
  \bibfield  {author} {\bibinfo {author} {\bibfnamefont {M.~V.}\ \bibnamefont
  {Larsen}}, \bibinfo {author} {\bibfnamefont {J.~E.}\ \bibnamefont
  {Bourassa}}, \bibinfo {author} {\bibfnamefont {S.}~\bibnamefont {Kocsis}},
  \bibinfo {author} {\bibfnamefont {J.~F.}\ \bibnamefont {Tasker}}, \bibinfo
  {author} {\bibfnamefont {R.~S.}\ \bibnamefont {Chadwick}}, \bibinfo {author}
  {\bibfnamefont {C.}~\bibnamefont {{Gonz{\'a}lez-Arciniegas}}}, \bibinfo
  {author} {\bibfnamefont {J.}~\bibnamefont {Hastrup}}, \bibinfo {author}
  {\bibfnamefont {C.~E.}\ \bibnamefont {{Lopetegui-Gonz{\'a}lez}}}, \bibinfo
  {author} {\bibfnamefont {F.~M.}\ \bibnamefont {Miatto}}, \bibinfo {author}
  {\bibfnamefont {A.}~\bibnamefont {Motamedi}}, \bibinfo {author}
  {\bibfnamefont {R.}~\bibnamefont {Noro}}, \bibinfo {author} {\bibfnamefont
  {G.}~\bibnamefont {Roeland}}, \bibinfo {author} {\bibfnamefont
  {R.}~\bibnamefont {Baby}}, \bibinfo {author} {\bibfnamefont {H.}~\bibnamefont
  {Chen}}, \bibinfo {author} {\bibfnamefont {P.}~\bibnamefont {Contu}},
  \bibinfo {author} {\bibfnamefont {I.}~\bibnamefont {Di~Luch}}, \bibinfo
  {author} {\bibfnamefont {C.}~\bibnamefont {Drago}}, \bibinfo {author}
  {\bibfnamefont {M.}~\bibnamefont {Giesbrecht}}, \bibinfo {author}
  {\bibfnamefont {T.}~\bibnamefont {Grainge}}, \bibinfo {author} {\bibfnamefont
  {I.}~\bibnamefont {Krasnokutska}}, \bibinfo {author} {\bibfnamefont
  {M.}~\bibnamefont {Menotti}}, \bibinfo {author} {\bibfnamefont
  {B.}~\bibnamefont {Morrison}}, \bibinfo {author} {\bibfnamefont
  {C.}~\bibnamefont {Puviraj}}, \bibinfo {author} {\bibfnamefont
  {K.}~\bibnamefont {Rezaei~Shad}}, \bibinfo {author} {\bibfnamefont
  {B.}~\bibnamefont {Hussain}}, \bibinfo {author} {\bibfnamefont
  {J.}~\bibnamefont {McMahon}}, \bibinfo {author} {\bibfnamefont {J.~E.}\
  \bibnamefont {Ortmann}}, \bibinfo {author} {\bibfnamefont {M.~J.}\
  \bibnamefont {Collins}}, \bibinfo {author} {\bibfnamefont {C.}~\bibnamefont
  {Ma}}, \bibinfo {author} {\bibfnamefont {D.~S.}\ \bibnamefont {Phillips}},
  \bibinfo {author} {\bibfnamefont {M.}~\bibnamefont {Seymour}}, \bibinfo
  {author} {\bibfnamefont {Q.~Y.}\ \bibnamefont {Tang}}, \bibinfo {author}
  {\bibfnamefont {B.}~\bibnamefont {Yang}}, \bibinfo {author} {\bibfnamefont
  {Z.}~\bibnamefont {Vernon}}, \bibinfo {author} {\bibfnamefont {R.~N.}\
  \bibnamefont {Alexander}},\ and\ \bibinfo {author} {\bibfnamefont {D.~H.}\
  \bibnamefont {Mahler}},\ }\bibfield  {title} {\bibinfo {title} {Integrated
  photonic source of {{Gottesman}}--{{Kitaev}}--{{Preskill}} qubits},\ }\href
  {https://doi.org/10.1038/s41586-025-09044-5} {\bibfield  {journal} {\bibinfo
  {journal} {Nature}\ }\textbf {\bibinfo {volume} {642}},\ \bibinfo {pages}
  {587} (\bibinfo {year} {2025})}\BibitemShut {NoStop}%
\bibitem [{\citenamefont {Ofek}\ \emph {et~al.}(2016)\citenamefont {Ofek},
  \citenamefont {Petrenko}, \citenamefont {Heeres}, \citenamefont {Reinhold},
  \citenamefont {Leghtas}, \citenamefont {Vlastakis}, \citenamefont {Liu},
  \citenamefont {Frunzio}, \citenamefont {Girvin}, \citenamefont {Jiang},
  \citenamefont {Mirrahimi}, \citenamefont {Devoret},\ and\ \citenamefont
  {Schoelkopf}}]{Ofek2016}%
  \BibitemOpen
  \bibfield  {author} {\bibinfo {author} {\bibfnamefont {N.}~\bibnamefont
  {Ofek}}, \bibinfo {author} {\bibfnamefont {A.}~\bibnamefont {Petrenko}},
  \bibinfo {author} {\bibfnamefont {R.}~\bibnamefont {Heeres}}, \bibinfo
  {author} {\bibfnamefont {P.}~\bibnamefont {Reinhold}}, \bibinfo {author}
  {\bibfnamefont {Z.}~\bibnamefont {Leghtas}}, \bibinfo {author} {\bibfnamefont
  {B.}~\bibnamefont {Vlastakis}}, \bibinfo {author} {\bibfnamefont
  {Y.}~\bibnamefont {Liu}}, \bibinfo {author} {\bibfnamefont {L.}~\bibnamefont
  {Frunzio}}, \bibinfo {author} {\bibfnamefont {S.~M.}\ \bibnamefont {Girvin}},
  \bibinfo {author} {\bibfnamefont {L.}~\bibnamefont {Jiang}}, \bibinfo
  {author} {\bibfnamefont {M.}~\bibnamefont {Mirrahimi}}, \bibinfo {author}
  {\bibfnamefont {M.~H.}\ \bibnamefont {Devoret}},\ and\ \bibinfo {author}
  {\bibfnamefont {R.~J.}\ \bibnamefont {Schoelkopf}},\ }\bibfield  {title}
  {\bibinfo {title} {Extending the lifetime of a quantum bit with error
  correction in superconducting circuits},\ }\href
  {https://doi.org/10.1038/nature18949} {\bibfield  {journal} {\bibinfo
  {journal} {Nature}\ }\textbf {\bibinfo {volume} {536}},\ \bibinfo {pages}
  {441} (\bibinfo {year} {2016})}\BibitemShut {NoStop}%
\bibitem [{\citenamefont {Sivak}\ \emph {et~al.}(2023)\citenamefont {Sivak},
  \citenamefont {Eickbusch}, \citenamefont {Royer}, \citenamefont {Singh},
  \citenamefont {Tsioutsios}, \citenamefont {Ganjam}, \citenamefont {Miano},
  \citenamefont {Brock}, \citenamefont {Ding}, \citenamefont {Frunzio},
  \citenamefont {Girvin}, \citenamefont {Schoelkopf},\ and\ \citenamefont
  {Devoret}}]{Sivak2023}%
  \BibitemOpen
  \bibfield  {author} {\bibinfo {author} {\bibfnamefont {V.~V.}\ \bibnamefont
  {Sivak}}, \bibinfo {author} {\bibfnamefont {A.}~\bibnamefont {Eickbusch}},
  \bibinfo {author} {\bibfnamefont {B.}~\bibnamefont {Royer}}, \bibinfo
  {author} {\bibfnamefont {S.}~\bibnamefont {Singh}}, \bibinfo {author}
  {\bibfnamefont {I.}~\bibnamefont {Tsioutsios}}, \bibinfo {author}
  {\bibfnamefont {S.}~\bibnamefont {Ganjam}}, \bibinfo {author} {\bibfnamefont
  {A.}~\bibnamefont {Miano}}, \bibinfo {author} {\bibfnamefont {B.~L.}\
  \bibnamefont {Brock}}, \bibinfo {author} {\bibfnamefont {A.~Z.}\ \bibnamefont
  {Ding}}, \bibinfo {author} {\bibfnamefont {L.}~\bibnamefont {Frunzio}},
  \bibinfo {author} {\bibfnamefont {S.~M.}\ \bibnamefont {Girvin}}, \bibinfo
  {author} {\bibfnamefont {R.~J.}\ \bibnamefont {Schoelkopf}},\ and\ \bibinfo
  {author} {\bibfnamefont {M.~H.}\ \bibnamefont {Devoret}},\ }\bibfield
  {title} {\bibinfo {title} {Real-time quantum error correction beyond
  break-even},\ }\href {https://doi.org/10.1038/s41586-023-05782-6} {\bibfield
  {journal} {\bibinfo  {journal} {Nature}\ }\textbf {\bibinfo {volume} {616}},\
  \bibinfo {pages} {50} (\bibinfo {year} {2023})}\BibitemShut {NoStop}%
\bibitem [{\citenamefont {Hastrup}\ and\ \citenamefont
  {Andersen}(2022)}]{Hastrup2022}%
  \BibitemOpen
  \bibfield  {author} {\bibinfo {author} {\bibfnamefont {J.}~\bibnamefont
  {Hastrup}}\ and\ \bibinfo {author} {\bibfnamefont {U.~L.}\ \bibnamefont
  {Andersen}},\ }\bibfield  {title} {\bibinfo {title} {All-optical cat-code
  quantum error correction},\ }\href
  {https://doi.org/10.1103/PhysRevResearch.4.043065} {\bibfield  {journal}
  {\bibinfo  {journal} {Phys. Rev. Research}\ }\textbf {\bibinfo {volume}
  {4}},\ \bibinfo {pages} {043065} (\bibinfo {year} {2022})}\BibitemShut
  {NoStop}%
\bibitem [{\citenamefont {Yin}\ \emph {et~al.}(2017)\citenamefont {Yin},
  \citenamefont {Cao}, \citenamefont {Li}, \citenamefont {Liao}, \citenamefont
  {Zhang}, \citenamefont {Ren}, \citenamefont {Cai}, \citenamefont {Liu},
  \citenamefont {Li}, \citenamefont {Dai}, \citenamefont {Li}, \citenamefont
  {Lu}, \citenamefont {Gong}, \citenamefont {Xu}, \citenamefont {Li},
  \citenamefont {Li}, \citenamefont {Yin}, \citenamefont {Jiang}, \citenamefont
  {Li}, \citenamefont {Jia}, \citenamefont {Ren}, \citenamefont {He},
  \citenamefont {Zhou}, \citenamefont {Zhang}, \citenamefont {Wang},
  \citenamefont {Chang}, \citenamefont {Zhu}, \citenamefont {Liu},
  \citenamefont {Chen}, \citenamefont {Lu}, \citenamefont {Shu}, \citenamefont
  {Peng}, \citenamefont {Wang},\ and\ \citenamefont {Pan}}]{Yin2017}%
  \BibitemOpen
  \bibfield  {author} {\bibinfo {author} {\bibfnamefont {J.}~\bibnamefont
  {Yin}}, \bibinfo {author} {\bibfnamefont {Y.}~\bibnamefont {Cao}}, \bibinfo
  {author} {\bibfnamefont {Y.-H.}\ \bibnamefont {Li}}, \bibinfo {author}
  {\bibfnamefont {S.-K.}\ \bibnamefont {Liao}}, \bibinfo {author}
  {\bibfnamefont {L.}~\bibnamefont {Zhang}}, \bibinfo {author} {\bibfnamefont
  {J.-G.}\ \bibnamefont {Ren}}, \bibinfo {author} {\bibfnamefont {W.-Q.}\
  \bibnamefont {Cai}}, \bibinfo {author} {\bibfnamefont {W.-Y.}\ \bibnamefont
  {Liu}}, \bibinfo {author} {\bibfnamefont {B.}~\bibnamefont {Li}}, \bibinfo
  {author} {\bibfnamefont {H.}~\bibnamefont {Dai}}, \bibinfo {author}
  {\bibfnamefont {G.-B.}\ \bibnamefont {Li}}, \bibinfo {author} {\bibfnamefont
  {Q.-M.}\ \bibnamefont {Lu}}, \bibinfo {author} {\bibfnamefont {Y.-H.}\
  \bibnamefont {Gong}}, \bibinfo {author} {\bibfnamefont {Y.}~\bibnamefont
  {Xu}}, \bibinfo {author} {\bibfnamefont {S.-L.}\ \bibnamefont {Li}}, \bibinfo
  {author} {\bibfnamefont {F.-Z.}\ \bibnamefont {Li}}, \bibinfo {author}
  {\bibfnamefont {Y.-Y.}\ \bibnamefont {Yin}}, \bibinfo {author} {\bibfnamefont
  {Z.-Q.}\ \bibnamefont {Jiang}}, \bibinfo {author} {\bibfnamefont
  {M.}~\bibnamefont {Li}}, \bibinfo {author} {\bibfnamefont {J.-J.}\
  \bibnamefont {Jia}}, \bibinfo {author} {\bibfnamefont {G.}~\bibnamefont
  {Ren}}, \bibinfo {author} {\bibfnamefont {D.}~\bibnamefont {He}}, \bibinfo
  {author} {\bibfnamefont {Y.-L.}\ \bibnamefont {Zhou}}, \bibinfo {author}
  {\bibfnamefont {X.-X.}\ \bibnamefont {Zhang}}, \bibinfo {author}
  {\bibfnamefont {N.}~\bibnamefont {Wang}}, \bibinfo {author} {\bibfnamefont
  {X.}~\bibnamefont {Chang}}, \bibinfo {author} {\bibfnamefont {Z.-C.}\
  \bibnamefont {Zhu}}, \bibinfo {author} {\bibfnamefont {N.-L.}\ \bibnamefont
  {Liu}}, \bibinfo {author} {\bibfnamefont {Y.-A.}\ \bibnamefont {Chen}},
  \bibinfo {author} {\bibfnamefont {C.-Y.}\ \bibnamefont {Lu}}, \bibinfo
  {author} {\bibfnamefont {R.}~\bibnamefont {Shu}}, \bibinfo {author}
  {\bibfnamefont {C.-Z.}\ \bibnamefont {Peng}}, \bibinfo {author}
  {\bibfnamefont {J.-Y.}\ \bibnamefont {Wang}},\ and\ \bibinfo {author}
  {\bibfnamefont {J.-W.}\ \bibnamefont {Pan}},\ }\bibfield  {title} {\bibinfo
  {title} {Satellite-based entanglement distribution over 1200 kilometers},\
  }\href {https://doi.org/10.1126/science.aan3211} {\bibfield  {journal}
  {\bibinfo  {journal} {Science}\ }\textbf {\bibinfo {volume} {356}},\ \bibinfo
  {pages} {1140} (\bibinfo {year} {2017})}\BibitemShut {NoStop}%
\bibitem [{\citenamefont {Wengerowsky}\ \emph {et~al.}(2020)\citenamefont
  {Wengerowsky}, \citenamefont {Joshi}, \citenamefont {Steinlechner},
  \citenamefont {Zichi}, \citenamefont {Liu}, \citenamefont {Scheidl},
  \citenamefont {Dobrovolskiy}, \citenamefont {{van der Molen}}, \citenamefont
  {Los}, \citenamefont {Zwiller}, \citenamefont {Versteegh}, \citenamefont
  {Mura}, \citenamefont {Calonico}, \citenamefont {Inguscio}, \citenamefont
  {Zeilinger}, \citenamefont {Xuereb},\ and\ \citenamefont
  {Ursin}}]{Wengerowsky2020}%
  \BibitemOpen
  \bibfield  {author} {\bibinfo {author} {\bibfnamefont {S.}~\bibnamefont
  {Wengerowsky}}, \bibinfo {author} {\bibfnamefont {S.~K.}\ \bibnamefont
  {Joshi}}, \bibinfo {author} {\bibfnamefont {F.}~\bibnamefont {Steinlechner}},
  \bibinfo {author} {\bibfnamefont {J.~R.}\ \bibnamefont {Zichi}}, \bibinfo
  {author} {\bibfnamefont {B.}~\bibnamefont {Liu}}, \bibinfo {author}
  {\bibfnamefont {T.}~\bibnamefont {Scheidl}}, \bibinfo {author} {\bibfnamefont
  {S.~M.}\ \bibnamefont {Dobrovolskiy}}, \bibinfo {author} {\bibfnamefont
  {R.}~\bibnamefont {{van der Molen}}}, \bibinfo {author} {\bibfnamefont
  {J.~W.}\ \bibnamefont {Los}}, \bibinfo {author} {\bibfnamefont
  {V.}~\bibnamefont {Zwiller}}, \bibinfo {author} {\bibfnamefont {M.~A.}\
  \bibnamefont {Versteegh}}, \bibinfo {author} {\bibfnamefont {A.}~\bibnamefont
  {Mura}}, \bibinfo {author} {\bibfnamefont {D.}~\bibnamefont {Calonico}},
  \bibinfo {author} {\bibfnamefont {M.}~\bibnamefont {Inguscio}}, \bibinfo
  {author} {\bibfnamefont {A.}~\bibnamefont {Zeilinger}}, \bibinfo {author}
  {\bibfnamefont {A.}~\bibnamefont {Xuereb}},\ and\ \bibinfo {author}
  {\bibfnamefont {R.}~\bibnamefont {Ursin}},\ }\bibfield  {title} {\bibinfo
  {title} {Passively stable distribution of polarisation entanglement over 192
  km of deployed optical fibre},\ }\href
  {https://doi.org/10.1038/s41534-019-0238-8} {\bibfield  {journal} {\bibinfo
  {journal} {npj Quantum Information}\ }\textbf {\bibinfo {volume} {6}},\
  \bibinfo {pages} {5} (\bibinfo {year} {2020})}\BibitemShut {NoStop}%
\bibitem [{\citenamefont {Neumann}\ \emph {et~al.}(2022)\citenamefont
  {Neumann}, \citenamefont {Buchner}, \citenamefont {Bulla}, \citenamefont
  {Bohmann},\ and\ \citenamefont {Ursin}}]{Neumann2022}%
  \BibitemOpen
  \bibfield  {author} {\bibinfo {author} {\bibfnamefont {S.~P.}\ \bibnamefont
  {Neumann}}, \bibinfo {author} {\bibfnamefont {A.}~\bibnamefont {Buchner}},
  \bibinfo {author} {\bibfnamefont {L.}~\bibnamefont {Bulla}}, \bibinfo
  {author} {\bibfnamefont {M.}~\bibnamefont {Bohmann}},\ and\ \bibinfo {author}
  {\bibfnamefont {R.}~\bibnamefont {Ursin}},\ }\bibfield  {title} {\bibinfo
  {title} {Continuous entanglement distribution over a transnational 248 km
  fiber link},\ }\href {https://doi.org/10.1038/s41467-022-33919-0} {\bibfield
  {journal} {\bibinfo  {journal} {Nat. Comm.}\ }\textbf {\bibinfo {volume}
  {13}},\ \bibinfo {pages} {6134} (\bibinfo {year} {2022})}\BibitemShut
  {NoStop}%
\bibitem [{\citenamefont {Gyger}\ \emph {et~al.}(2022)\citenamefont {Gyger},
  \citenamefont {Zeuner}, \citenamefont {Lettner}, \citenamefont {Bensoussan},
  \citenamefont {Carln{\"a}s}, \citenamefont {Ekemar}, \citenamefont
  {Schweickert}, \citenamefont {Hedlund}, \citenamefont {Hammar}, \citenamefont
  {Nilsson}, \citenamefont {Alml{\"o}f}, \citenamefont {Steinhauer},
  \citenamefont {Llosera},\ and\ \citenamefont {Zwiller}}]{Gyger2022}%
  \BibitemOpen
  \bibfield  {author} {\bibinfo {author} {\bibfnamefont {S.}~\bibnamefont
  {Gyger}}, \bibinfo {author} {\bibfnamefont {K.~D.}\ \bibnamefont {Zeuner}},
  \bibinfo {author} {\bibfnamefont {T.}~\bibnamefont {Lettner}}, \bibinfo
  {author} {\bibfnamefont {S.}~\bibnamefont {Bensoussan}}, \bibinfo {author}
  {\bibfnamefont {M.}~\bibnamefont {Carln{\"a}s}}, \bibinfo {author}
  {\bibfnamefont {L.}~\bibnamefont {Ekemar}}, \bibinfo {author} {\bibfnamefont
  {L.}~\bibnamefont {Schweickert}}, \bibinfo {author} {\bibfnamefont {C.~R.}\
  \bibnamefont {Hedlund}}, \bibinfo {author} {\bibfnamefont {M.}~\bibnamefont
  {Hammar}}, \bibinfo {author} {\bibfnamefont {T.}~\bibnamefont {Nilsson}},
  \bibinfo {author} {\bibfnamefont {J.}~\bibnamefont {Alml{\"o}f}}, \bibinfo
  {author} {\bibfnamefont {S.}~\bibnamefont {Steinhauer}}, \bibinfo {author}
  {\bibfnamefont {G.~V.}\ \bibnamefont {Llosera}},\ and\ \bibinfo {author}
  {\bibfnamefont {V.}~\bibnamefont {Zwiller}},\ }\bibfield  {title} {\bibinfo
  {title} {Metropolitan single-photon distribution at 1550 nm for random number
  generation},\ }\href {https://doi.org/10.1063/5.0112939} {\bibfield
  {journal} {\bibinfo  {journal} {Appl. Phys. Lett.}\ }\textbf {\bibinfo
  {volume} {121}},\ \bibinfo {pages} {194003} (\bibinfo {year}
  {2022})}\BibitemShut {NoStop}%
\bibitem [{\citenamefont {Peuntinger}\ \emph {et~al.}(2014)\citenamefont
  {Peuntinger}, \citenamefont {Heim}, \citenamefont {M{\"u}ller}, \citenamefont
  {Gabriel}, \citenamefont {Marquardt},\ and\ \citenamefont
  {Leuchs}}]{Peuntinger2014}%
  \BibitemOpen
  \bibfield  {author} {\bibinfo {author} {\bibfnamefont {C.}~\bibnamefont
  {Peuntinger}}, \bibinfo {author} {\bibfnamefont {B.}~\bibnamefont {Heim}},
  \bibinfo {author} {\bibfnamefont {C.~R.}\ \bibnamefont {M{\"u}ller}},
  \bibinfo {author} {\bibfnamefont {C.}~\bibnamefont {Gabriel}}, \bibinfo
  {author} {\bibfnamefont {C.}~\bibnamefont {Marquardt}},\ and\ \bibinfo
  {author} {\bibfnamefont {G.}~\bibnamefont {Leuchs}},\ }\bibfield  {title}
  {\bibinfo {title} {Distribution of squeezed states through an atmospheric
  channel},\ }\href {https://doi.org/10.1103/PhysRevLett.113.060502} {\bibfield
   {journal} {\bibinfo  {journal} {Phys. Rev. Lett.}\ }\textbf {\bibinfo
  {volume} {113}},\ \bibinfo {pages} {060502} (\bibinfo {year}
  {2014})}\BibitemShut {NoStop}%
\bibitem [{\citenamefont {Suleiman}\ \emph {et~al.}(2022)\citenamefont
  {Suleiman}, \citenamefont {Nielsen}, \citenamefont {Guo}, \citenamefont
  {Jain}, \citenamefont {{Neergaard-Nielsen}}, \citenamefont {Gehring},\ and\
  \citenamefont {Andersen}}]{Suleiman2022}%
  \BibitemOpen
  \bibfield  {author} {\bibinfo {author} {\bibfnamefont {I.}~\bibnamefont
  {Suleiman}}, \bibinfo {author} {\bibfnamefont {J.~A.}\ \bibnamefont
  {Nielsen}}, \bibinfo {author} {\bibfnamefont {X.}~\bibnamefont {Guo}},
  \bibinfo {author} {\bibfnamefont {N.}~\bibnamefont {Jain}}, \bibinfo {author}
  {\bibfnamefont {J.}~\bibnamefont {{Neergaard-Nielsen}}}, \bibinfo {author}
  {\bibfnamefont {T.}~\bibnamefont {Gehring}},\ and\ \bibinfo {author}
  {\bibfnamefont {U.~L.}\ \bibnamefont {Andersen}},\ }\bibfield  {title}
  {\bibinfo {title} {40 km fiber transmission of squeezed light measured with a
  real local oscillator},\ }\href {https://doi.org/10.1088/2058-9565/ac7ba1}
  {\bibfield  {journal} {\bibinfo  {journal} {Quantum Science and Technology}\
  }\textbf {\bibinfo {volume} {7}},\ \bibinfo {pages} {045003} (\bibinfo {year}
  {2022})}\BibitemShut {NoStop}%
\bibitem [{\citenamefont {Dakna}\ \emph {et~al.}(1997)\citenamefont {Dakna},
  \citenamefont {Anhut}, \citenamefont {Opatrn{\'y}}, \citenamefont
  {Kn{\"o}ll},\ and\ \citenamefont {Welsch}}]{Dakna1997}%
  \BibitemOpen
  \bibfield  {author} {\bibinfo {author} {\bibfnamefont {M.}~\bibnamefont
  {Dakna}}, \bibinfo {author} {\bibfnamefont {T.}~\bibnamefont {Anhut}},
  \bibinfo {author} {\bibfnamefont {T.}~\bibnamefont {Opatrn{\'y}}}, \bibinfo
  {author} {\bibfnamefont {L.}~\bibnamefont {Kn{\"o}ll}},\ and\ \bibinfo
  {author} {\bibfnamefont {D.-G.}\ \bibnamefont {Welsch}},\ }\bibfield  {title}
  {\bibinfo {title} {Generating {{Schr{\"o}dinger-cat-like}} states by means of
  conditional measurements on a beam splitter},\ }\href
  {https://doi.org/10.1103/PhysRevA.55.3184} {\bibfield  {journal} {\bibinfo
  {journal} {Phys. Rev. A}\ }\textbf {\bibinfo {volume} {55}},\ \bibinfo
  {pages} {3184} (\bibinfo {year} {1997})}\BibitemShut {NoStop}%
\bibitem [{\citenamefont {Lvovsky}\ \emph {et~al.}(2020)\citenamefont
  {Lvovsky}, \citenamefont {Grangier}, \citenamefont {Ourjoumtsev},
  \citenamefont {Parigi}, \citenamefont {Sasaki},\ and\ \citenamefont
  {{Tualle-Brouri}}}]{Lvovsky2020-ProductionApplicationsNonGaussiana}%
  \BibitemOpen
  \bibfield  {author} {\bibinfo {author} {\bibfnamefont {A.~I.}\ \bibnamefont
  {Lvovsky}}, \bibinfo {author} {\bibfnamefont {P.}~\bibnamefont {Grangier}},
  \bibinfo {author} {\bibfnamefont {A.}~\bibnamefont {Ourjoumtsev}}, \bibinfo
  {author} {\bibfnamefont {V.}~\bibnamefont {Parigi}}, \bibinfo {author}
  {\bibfnamefont {M.}~\bibnamefont {Sasaki}},\ and\ \bibinfo {author}
  {\bibfnamefont {R.}~\bibnamefont {{Tualle-Brouri}}},\ }\bibfield  {title}
  {\bibinfo {title} {Production and applications of non-{{Gaussian}} quantum
  states of light},\ }\href@noop {} {\bibfield  {journal} {\bibinfo  {journal}
  {arXiv:2006.16985}\ } (\bibinfo {year} {2020})}\BibitemShut {NoStop}%
\bibitem [{\citenamefont {Tipsmark}\ \emph {et~al.}(2011)\citenamefont
  {Tipsmark}, \citenamefont {Dong}, \citenamefont {Laghaout}, \citenamefont
  {Marek}, \citenamefont {Jezek},\ and\ \citenamefont
  {Andersen}}]{Tipsmark2011}%
  \BibitemOpen
  \bibfield  {author} {\bibinfo {author} {\bibfnamefont {A.}~\bibnamefont
  {Tipsmark}}, \bibinfo {author} {\bibfnamefont {R.}~\bibnamefont {Dong}},
  \bibinfo {author} {\bibfnamefont {A.}~\bibnamefont {Laghaout}}, \bibinfo
  {author} {\bibfnamefont {P.}~\bibnamefont {Marek}}, \bibinfo {author}
  {\bibfnamefont {M.}~\bibnamefont {Jezek}},\ and\ \bibinfo {author}
  {\bibfnamefont {U.~L.}\ \bibnamefont {Andersen}},\ }\bibfield  {title}
  {\bibinfo {title} {Experimental demonstration of a {{Hadamard}} gate for
  coherent state qubits},\ }\href {https://doi.org/10.1103/physreva.84.050301}
  {\bibfield  {journal} {\bibinfo  {journal} {Phys. Rev. A}\ }\textbf {\bibinfo
  {volume} {84}},\ \bibinfo {pages} {050301} (\bibinfo {year}
  {2011})}\BibitemShut {NoStop}%
\bibitem [{\citenamefont {Blandino}\ \emph {et~al.}(2012)\citenamefont
  {Blandino}, \citenamefont {Ferreyrol}, \citenamefont {Barbieri},
  \citenamefont {Grangier},\ and\ \citenamefont
  {{Tualle-Brouri}}}]{Blandino2012a}%
  \BibitemOpen
  \bibfield  {author} {\bibinfo {author} {\bibfnamefont {R.}~\bibnamefont
  {Blandino}}, \bibinfo {author} {\bibfnamefont {F.}~\bibnamefont {Ferreyrol}},
  \bibinfo {author} {\bibfnamefont {M.}~\bibnamefont {Barbieri}}, \bibinfo
  {author} {\bibfnamefont {P.}~\bibnamefont {Grangier}},\ and\ \bibinfo
  {author} {\bibfnamefont {R.}~\bibnamefont {{Tualle-Brouri}}},\ }\bibfield
  {title} {\bibinfo {title} {Characterization of a {{$\pi$}}-phase shift
  quantum gate for coherent-state qubits},\ }\href
  {https://doi.org/10.1088/1367-2630/14/1/013017} {\bibfield  {journal}
  {\bibinfo  {journal} {New Journal of Physics}\ }\textbf {\bibinfo {volume}
  {14}},\ \bibinfo {pages} {013017} (\bibinfo {year} {2012})}\BibitemShut
  {NoStop}%
\bibitem [{\citenamefont {Takahashi}\ \emph {et~al.}(2010)\citenamefont
  {Takahashi}, \citenamefont {Neergaard-Nielsen}, \citenamefont {Takeuchi},
  \citenamefont {Takeoka}, \citenamefont {Hayasaka}, \citenamefont {Furusawa},\
  and\ \citenamefont {Sasaki}}]{takahashi_entanglement_2010}%
  \BibitemOpen
  \bibfield  {author} {\bibinfo {author} {\bibfnamefont {H.}~\bibnamefont
  {Takahashi}}, \bibinfo {author} {\bibfnamefont {J.~S.}\ \bibnamefont
  {Neergaard-Nielsen}}, \bibinfo {author} {\bibfnamefont {M.}~\bibnamefont
  {Takeuchi}}, \bibinfo {author} {\bibfnamefont {M.}~\bibnamefont {Takeoka}},
  \bibinfo {author} {\bibfnamefont {K.}~\bibnamefont {Hayasaka}}, \bibinfo
  {author} {\bibfnamefont {A.}~\bibnamefont {Furusawa}},\ and\ \bibinfo
  {author} {\bibfnamefont {M.}~\bibnamefont {Sasaki}},\ }\bibfield  {title}
  {\bibinfo {title} {Entanglement distillation from {Gaussian} input states},\
  }\href {https://doi.org/10.1038/nphoton.2010.1} {\bibfield  {journal}
  {\bibinfo  {journal} {Nat. Phot.}\ }\textbf {\bibinfo {volume} {4}},\
  \bibinfo {pages} {178} (\bibinfo {year} {2010})}\BibitemShut {NoStop}%
\bibitem [{\citenamefont {Parigi}\ \emph {et~al.}(2007)\citenamefont {Parigi},
  \citenamefont {Zavatta}, \citenamefont {Kim},\ and\ \citenamefont
  {Bellini}}]{parigi_probing_2007}%
  \BibitemOpen
  \bibfield  {author} {\bibinfo {author} {\bibfnamefont {V.}~\bibnamefont
  {Parigi}}, \bibinfo {author} {\bibfnamefont {A.}~\bibnamefont {Zavatta}},
  \bibinfo {author} {\bibfnamefont {M.~S.}\ \bibnamefont {Kim}},\ and\ \bibinfo
  {author} {\bibfnamefont {M.}~\bibnamefont {Bellini}},\ }\bibfield  {title}
  {\bibinfo {title} {Probing quantum commutation rules by addition and
  subtraction of single photons to/from a light field.},\ }\href
  {https://doi.org/10.1126/science.1146204} {\bibfield  {journal} {\bibinfo
  {journal} {Science}\ }\textbf {\bibinfo {volume} {317}},\ \bibinfo {pages}
  {1890} (\bibinfo {year} {2007})}\BibitemShut {NoStop}%
\bibitem [{\citenamefont {Ourjoumtsev}\ \emph {et~al.}(2009)\citenamefont
  {Ourjoumtsev}, \citenamefont {Ferreyrol}, \citenamefont {{Tualle-Brouri}},\
  and\ \citenamefont {Grangier}}]{Ourjoumtsev2009}%
  \BibitemOpen
  \bibfield  {author} {\bibinfo {author} {\bibfnamefont {A.}~\bibnamefont
  {Ourjoumtsev}}, \bibinfo {author} {\bibfnamefont {F.}~\bibnamefont
  {Ferreyrol}}, \bibinfo {author} {\bibfnamefont {R.}~\bibnamefont
  {{Tualle-Brouri}}},\ and\ \bibinfo {author} {\bibfnamefont {P.}~\bibnamefont
  {Grangier}},\ }\bibfield  {title} {\bibinfo {title} {Preparation of non-local
  superpositions of quasi-classical light states},\ }\href
  {https://doi.org/10.1038/nphys1199} {\bibfield  {journal} {\bibinfo
  {journal} {Nature Physics}\ }\textbf {\bibinfo {volume} {5}},\ \bibinfo
  {pages} {189} (\bibinfo {year} {2009})}\BibitemShut {NoStop}%
\bibitem [{\citenamefont {Jeong}\ \emph {et~al.}(2014)\citenamefont {Jeong},
  \citenamefont {Zavatta}, \citenamefont {Kang}, \citenamefont {Lee},
  \citenamefont {Costanzo}, \citenamefont {Grandi}, \citenamefont {Ralph},\
  and\ \citenamefont {Bellini}}]{jeong_generation_2014}%
  \BibitemOpen
  \bibfield  {author} {\bibinfo {author} {\bibfnamefont {H.}~\bibnamefont
  {Jeong}}, \bibinfo {author} {\bibfnamefont {A.}~\bibnamefont {Zavatta}},
  \bibinfo {author} {\bibfnamefont {M.}~\bibnamefont {Kang}}, \bibinfo {author}
  {\bibfnamefont {S.-w.}\ \bibnamefont {Lee}}, \bibinfo {author} {\bibfnamefont
  {L.~S.}\ \bibnamefont {Costanzo}}, \bibinfo {author} {\bibfnamefont
  {S.}~\bibnamefont {Grandi}}, \bibinfo {author} {\bibfnamefont {T.~C.}\
  \bibnamefont {Ralph}},\ and\ \bibinfo {author} {\bibfnamefont
  {M.}~\bibnamefont {Bellini}},\ }\bibfield  {title} {\bibinfo {title}
  {Generation of hybrid entanglement of light},\ }\href
  {https://doi.org/10.1038/nphoton.2014.136} {\bibfield  {journal} {\bibinfo
  {journal} {Nat. Phot.}\ }\textbf {\bibinfo {volume} {8}},\ \bibinfo {pages}
  {564} (\bibinfo {year} {2014})}\BibitemShut {NoStop}%
\bibitem [{\citenamefont {Morin}\ \emph {et~al.}(2014)\citenamefont {Morin},
  \citenamefont {Huang}, \citenamefont {Liu}, \citenamefont {Jeannic},
  \citenamefont {Fabre},\ and\ \citenamefont {Laurat}}]{morin_remote_2014}%
  \BibitemOpen
  \bibfield  {author} {\bibinfo {author} {\bibfnamefont {O.}~\bibnamefont
  {Morin}}, \bibinfo {author} {\bibfnamefont {K.}~\bibnamefont {Huang}},
  \bibinfo {author} {\bibfnamefont {J.}~\bibnamefont {Liu}}, \bibinfo {author}
  {\bibfnamefont {H.~L.}\ \bibnamefont {Jeannic}}, \bibinfo {author}
  {\bibfnamefont {C.}~\bibnamefont {Fabre}},\ and\ \bibinfo {author}
  {\bibfnamefont {J.}~\bibnamefont {Laurat}},\ }\bibfield  {title} {\bibinfo
  {title} {Remote creation of hybrid entanglement between particle-like and
  wave-like optical qubits},\ }\href {https://doi.org/10.1038/nphoton.2014.137}
  {\bibfield  {journal} {\bibinfo  {journal} {Nat. Phot.}\ }\textbf {\bibinfo
  {volume} {8}},\ \bibinfo {pages} {570} (\bibinfo {year} {2014})}\BibitemShut
  {NoStop}%
\bibitem [{\citenamefont {{Neergaard-Nielsen}}\ \emph
  {et~al.}(2010)\citenamefont {{Neergaard-Nielsen}}, \citenamefont {Takeuchi},
  \citenamefont {Wakui}, \citenamefont {Takahashi}, \citenamefont {Hayasaka},
  \citenamefont {Takeoka},\ and\ \citenamefont
  {Sasaki}}]{Neergaard-Nielsen2010}%
  \BibitemOpen
  \bibfield  {author} {\bibinfo {author} {\bibfnamefont {J.~S.}\ \bibnamefont
  {{Neergaard-Nielsen}}}, \bibinfo {author} {\bibfnamefont {M.}~\bibnamefont
  {Takeuchi}}, \bibinfo {author} {\bibfnamefont {K.}~\bibnamefont {Wakui}},
  \bibinfo {author} {\bibfnamefont {H.}~\bibnamefont {Takahashi}}, \bibinfo
  {author} {\bibfnamefont {K.}~\bibnamefont {Hayasaka}}, \bibinfo {author}
  {\bibfnamefont {M.}~\bibnamefont {Takeoka}},\ and\ \bibinfo {author}
  {\bibfnamefont {M.}~\bibnamefont {Sasaki}},\ }\bibfield  {title} {\bibinfo
  {title} {Optical continuous-variable qubit},\ }\href
  {https://doi.org/10.1103/PhysRevLett.105.053602} {\bibfield  {journal}
  {\bibinfo  {journal} {Phys. Rev. Lett.}\ }\textbf {\bibinfo {volume} {105}},\
  \bibinfo {pages} {053602} (\bibinfo {year} {2010})}\BibitemShut {NoStop}%
\bibitem [{\citenamefont {Neergaard-Nielsen}\ \emph {et~al.}(2013)\citenamefont
  {Neergaard-Nielsen}, \citenamefont {Eto}, \citenamefont {Lee}, \citenamefont
  {Jeong},\ and\ \citenamefont {Sasaki}}]{neergaard-nielsen_quantum_2013}%
  \BibitemOpen
  \bibfield  {author} {\bibinfo {author} {\bibfnamefont {J.~S.}\ \bibnamefont
  {Neergaard-Nielsen}}, \bibinfo {author} {\bibfnamefont {Y.}~\bibnamefont
  {Eto}}, \bibinfo {author} {\bibfnamefont {C.-W.}\ \bibnamefont {Lee}},
  \bibinfo {author} {\bibfnamefont {H.}~\bibnamefont {Jeong}},\ and\ \bibinfo
  {author} {\bibfnamefont {M.}~\bibnamefont {Sasaki}},\ }\bibfield  {title}
  {\bibinfo {title} {Quantum tele-amplification with a continuous-variable
  superposition state},\ }\href {https://doi.org/10.1038/nphoton.2013.101}
  {\bibfield  {journal} {\bibinfo  {journal} {Nat. Phot.}\ }\textbf {\bibinfo
  {volume} {7}},\ \bibinfo {pages} {439} (\bibinfo {year} {2013})}\BibitemShut
  {NoStop}%
\bibitem [{\citenamefont {Lee}\ \emph {et~al.}(2011)\citenamefont {Lee},
  \citenamefont {Benichi}, \citenamefont {Takeno}, \citenamefont {Takeda},
  \citenamefont {Webb}, \citenamefont {Huntington},\ and\ \citenamefont
  {Furusawa}}]{Lee2011g}%
  \BibitemOpen
  \bibfield  {author} {\bibinfo {author} {\bibfnamefont {N.}~\bibnamefont
  {Lee}}, \bibinfo {author} {\bibfnamefont {H.}~\bibnamefont {Benichi}},
  \bibinfo {author} {\bibfnamefont {Y.}~\bibnamefont {Takeno}}, \bibinfo
  {author} {\bibfnamefont {S.}~\bibnamefont {Takeda}}, \bibinfo {author}
  {\bibfnamefont {J.}~\bibnamefont {Webb}}, \bibinfo {author} {\bibfnamefont
  {E.}~\bibnamefont {Huntington}},\ and\ \bibinfo {author} {\bibfnamefont
  {A.}~\bibnamefont {Furusawa}},\ }\bibfield  {title} {\bibinfo {title}
  {Teleportation of nonclassical wave packets of light.},\ }\href
  {https://doi.org/10.1126/science.1201034} {\bibfield  {journal} {\bibinfo
  {journal} {Science}\ }\textbf {\bibinfo {volume} {332}},\ \bibinfo {pages}
  {330} (\bibinfo {year} {2011})}\BibitemShut {NoStop}%
\bibitem [{\citenamefont {Sychev}\ \emph {et~al.}(2017)\citenamefont {Sychev},
  \citenamefont {Ulanov}, \citenamefont {Pushkina}, \citenamefont {Richards},
  \citenamefont {Fedorov},\ and\ \citenamefont {Lvovsky}}]{Sychev2017}%
  \BibitemOpen
  \bibfield  {author} {\bibinfo {author} {\bibfnamefont {D.~V.}\ \bibnamefont
  {Sychev}}, \bibinfo {author} {\bibfnamefont {A.~E.}\ \bibnamefont {Ulanov}},
  \bibinfo {author} {\bibfnamefont {A.~A.}\ \bibnamefont {Pushkina}}, \bibinfo
  {author} {\bibfnamefont {M.~W.}\ \bibnamefont {Richards}}, \bibinfo {author}
  {\bibfnamefont {I.~A.}\ \bibnamefont {Fedorov}},\ and\ \bibinfo {author}
  {\bibfnamefont {A.~I.}\ \bibnamefont {Lvovsky}},\ }\bibfield  {title}
  {\bibinfo {title} {Enlargement of optical {{Schr{\"o}dinger}}'s cat states},\
  }\href {https://doi.org/10.1038/nphoton.2017.57} {\bibfield  {journal}
  {\bibinfo  {journal} {Nat. Phot.}\ }\textbf {\bibinfo {volume} {11}},\
  \bibinfo {pages} {379} (\bibinfo {year} {2017})}\BibitemShut {NoStop}%
\bibitem [{\citenamefont {Konno}\ \emph {et~al.}(2024)\citenamefont {Konno},
  \citenamefont {Asavanant}, \citenamefont {Hanamura}, \citenamefont
  {Nagayoshi}, \citenamefont {Fukui}, \citenamefont {Sakaguchi}, \citenamefont
  {Ide}, \citenamefont {China}, \citenamefont {Yabuno}, \citenamefont {Miki},
  \citenamefont {Terai}, \citenamefont {Takase}, \citenamefont {Endo},
  \citenamefont {Marek}, \citenamefont {Filip}, \citenamefont {{van Loock}},\
  and\ \citenamefont {Furusawa}}]{Konno2024}%
  \BibitemOpen
  \bibfield  {author} {\bibinfo {author} {\bibfnamefont {S.}~\bibnamefont
  {Konno}}, \bibinfo {author} {\bibfnamefont {W.}~\bibnamefont {Asavanant}},
  \bibinfo {author} {\bibfnamefont {F.}~\bibnamefont {Hanamura}}, \bibinfo
  {author} {\bibfnamefont {H.}~\bibnamefont {Nagayoshi}}, \bibinfo {author}
  {\bibfnamefont {K.}~\bibnamefont {Fukui}}, \bibinfo {author} {\bibfnamefont
  {A.}~\bibnamefont {Sakaguchi}}, \bibinfo {author} {\bibfnamefont
  {R.}~\bibnamefont {Ide}}, \bibinfo {author} {\bibfnamefont {F.}~\bibnamefont
  {China}}, \bibinfo {author} {\bibfnamefont {M.}~\bibnamefont {Yabuno}},
  \bibinfo {author} {\bibfnamefont {S.}~\bibnamefont {Miki}}, \bibinfo {author}
  {\bibfnamefont {H.}~\bibnamefont {Terai}}, \bibinfo {author} {\bibfnamefont
  {K.}~\bibnamefont {Takase}}, \bibinfo {author} {\bibfnamefont
  {M.}~\bibnamefont {Endo}}, \bibinfo {author} {\bibfnamefont {P.}~\bibnamefont
  {Marek}}, \bibinfo {author} {\bibfnamefont {R.}~\bibnamefont {Filip}},
  \bibinfo {author} {\bibfnamefont {P.}~\bibnamefont {{van Loock}}},\ and\
  \bibinfo {author} {\bibfnamefont {A.}~\bibnamefont {Furusawa}},\ }\bibfield
  {title} {\bibinfo {title} {Propagating {{Gottesman-Kitaev-Preskill}} states
  encoded in an optical oscillator},\ }\href
  {https://doi.org/10.1126/science.adk7560} {\bibfield  {journal} {\bibinfo
  {journal} {Science}\ }\textbf {\bibinfo {volume} {383}},\ \bibinfo {pages}
  {289} (\bibinfo {year} {2024})}\BibitemShut {NoStop}%
\bibitem [{\citenamefont {Wang}\ \emph {et~al.}(2022)\citenamefont {Wang},
  \citenamefont {Zhang}, \citenamefont {Qin}, \citenamefont {Zhang},
  \citenamefont {Zeng}, \citenamefont {Su}, \citenamefont {Xie},\ and\
  \citenamefont {Peng}}]{Wang2022c}%
  \BibitemOpen
  \bibfield  {author} {\bibinfo {author} {\bibfnamefont {M.}~\bibnamefont
  {Wang}}, \bibinfo {author} {\bibfnamefont {M.}~\bibnamefont {Zhang}},
  \bibinfo {author} {\bibfnamefont {Z.}~\bibnamefont {Qin}}, \bibinfo {author}
  {\bibfnamefont {Q.}~\bibnamefont {Zhang}}, \bibinfo {author} {\bibfnamefont
  {L.}~\bibnamefont {Zeng}}, \bibinfo {author} {\bibfnamefont {X.}~\bibnamefont
  {Su}}, \bibinfo {author} {\bibfnamefont {C.}~\bibnamefont {Xie}},\ and\
  \bibinfo {author} {\bibfnamefont {K.}~\bibnamefont {Peng}},\ }\bibfield
  {title} {\bibinfo {title} {Experimental preparation and manipulation of
  squeezed cat states via an all-optical in-line squeezer},\ }\href
  {https://doi.org/10.1002/lpor.202200336} {\bibfield  {journal} {\bibinfo
  {journal} {Laser \& Photonics Reviews}\ }\textbf {\bibinfo {volume} {16}},\
  \bibinfo {pages} {2200336} (\bibinfo {year} {2022})}\BibitemShut {NoStop}%
\bibitem [{\citenamefont {Yoshida}\ \emph {et~al.}(2025)\citenamefont
  {Yoshida}, \citenamefont {Okuno}, \citenamefont {Kashiwazaki}, \citenamefont
  {Umeki}, \citenamefont {Miki}, \citenamefont {China}, \citenamefont {Yabuno},
  \citenamefont {Terai},\ and\ \citenamefont {Takeda}}]{Yoshida2025}%
  \BibitemOpen
  \bibfield  {author} {\bibinfo {author} {\bibfnamefont {T.}~\bibnamefont
  {Yoshida}}, \bibinfo {author} {\bibfnamefont {D.}~\bibnamefont {Okuno}},
  \bibinfo {author} {\bibfnamefont {T.}~\bibnamefont {Kashiwazaki}}, \bibinfo
  {author} {\bibfnamefont {T.}~\bibnamefont {Umeki}}, \bibinfo {author}
  {\bibfnamefont {S.}~\bibnamefont {Miki}}, \bibinfo {author} {\bibfnamefont
  {F.}~\bibnamefont {China}}, \bibinfo {author} {\bibfnamefont
  {M.}~\bibnamefont {Yabuno}}, \bibinfo {author} {\bibfnamefont
  {H.}~\bibnamefont {Terai}},\ and\ \bibinfo {author} {\bibfnamefont
  {S.}~\bibnamefont {Takeda}},\ }\bibfield  {title} {\bibinfo {title}
  {Sequential and {{Programmable Squeezing Gates}} for {{Optical Non-Gaussian
  Input States}}},\ }\href {https://doi.org/10.1103/PRXQuantum.6.010311}
  {\bibfield  {journal} {\bibinfo  {journal} {PRX Quantum}\ }\textbf {\bibinfo
  {volume} {6}},\ \bibinfo {pages} {010311} (\bibinfo {year}
  {2025})}\BibitemShut {NoStop}%
\bibitem [{\citenamefont {Grebien}\ \emph {et~al.}(2022)\citenamefont
  {Grebien}, \citenamefont {Goettsch}, \citenamefont {Hage}, \citenamefont
  {Fiurasek},\ and\ \citenamefont {Schnabel}}]{Grebien2022}%
  \BibitemOpen
  \bibfield  {author} {\bibinfo {author} {\bibfnamefont {S.}~\bibnamefont
  {Grebien}}, \bibinfo {author} {\bibfnamefont {J.}~\bibnamefont {Goettsch}},
  \bibinfo {author} {\bibfnamefont {B.}~\bibnamefont {Hage}}, \bibinfo {author}
  {\bibfnamefont {J.}~\bibnamefont {Fiurasek}},\ and\ \bibinfo {author}
  {\bibfnamefont {R.}~\bibnamefont {Schnabel}},\ }\bibfield  {title} {\bibinfo
  {title} {Multi-step two-copy distillation of squeezed states via two photon
  subtraction},\ }\href {https://doi.org/10.1103/PhysRevLett.129.273604}
  {\bibfield  {journal} {\bibinfo  {journal} {Phys. Rev. Lett.}\ }\textbf
  {\bibinfo {volume} {129}},\ \bibinfo {pages} {273604} (\bibinfo {year}
  {2022})}\BibitemShut {NoStop}%
\bibitem [{\citenamefont {Baune}\ \emph {et~al.}(2017)\citenamefont {Baune},
  \citenamefont {Fiur{\'a}{\v s}ek},\ and\ \citenamefont
  {Schnabel}}]{Baune2017}%
  \BibitemOpen
  \bibfield  {author} {\bibinfo {author} {\bibfnamefont {C.}~\bibnamefont
  {Baune}}, \bibinfo {author} {\bibfnamefont {J.}~\bibnamefont {Fiur{\'a}{\v
  s}ek}},\ and\ \bibinfo {author} {\bibfnamefont {R.}~\bibnamefont
  {Schnabel}},\ }\bibfield  {title} {\bibinfo {title} {Negative {{Wigner}}
  function at telecommunication wavelength from homodyne detection},\ }\href
  {https://doi.org/10.1103/PhysRevA.95.061802} {\bibfield  {journal} {\bibinfo
  {journal} {Phys. Rev. A}\ }\textbf {\bibinfo {volume} {95}},\ \bibinfo
  {pages} {061802} (\bibinfo {year} {2017})}\BibitemShut {NoStop}%
\bibitem [{\citenamefont {Namekata}\ \emph {et~al.}(2010)\citenamefont
  {Namekata}, \citenamefont {Takahashi}, \citenamefont {Fujii}, \citenamefont
  {Fukuda}, \citenamefont {Kurimura},\ and\ \citenamefont
  {Inoue}}]{Namekata2010}%
  \BibitemOpen
  \bibfield  {author} {\bibinfo {author} {\bibfnamefont {N.}~\bibnamefont
  {Namekata}}, \bibinfo {author} {\bibfnamefont {Y.}~\bibnamefont {Takahashi}},
  \bibinfo {author} {\bibfnamefont {G.}~\bibnamefont {Fujii}}, \bibinfo
  {author} {\bibfnamefont {D.}~\bibnamefont {Fukuda}}, \bibinfo {author}
  {\bibfnamefont {S.}~\bibnamefont {Kurimura}},\ and\ \bibinfo {author}
  {\bibfnamefont {S.}~\bibnamefont {Inoue}},\ }\bibfield  {title} {\bibinfo
  {title} {Non-{{Gaussian}} operation based on photon subtraction using a
  photon-number-resolving detector at a telecommunications wavelength},\ }\href
  {https://doi.org/10.1038/nphoton.2010.158} {\bibfield  {journal} {\bibinfo
  {journal} {Nat. Phot.}\ }\textbf {\bibinfo {volume} {4}},\ \bibinfo {pages}
  {655} (\bibinfo {year} {2010})}\BibitemShut {NoStop}%
\bibitem [{\citenamefont {Breum}(2020)}]{Breum2020}%
  \BibitemOpen
  \bibfield  {author} {\bibinfo {author} {\bibfnamefont {C.~R.}\ \bibnamefont
  {Breum}},\ }\emph {\bibinfo {title} {Quantum {{Communication}} with
  Non-{{Gaussian}} States}},\ \href
  {https://orbit.dtu.dk/en/publications/quantum-communication-with-non-gaussian-states}
  {Ph.D. thesis},\ \bibinfo  {school} {Technical University of Denmark}
  (\bibinfo {year} {2020})\BibitemShut {NoStop}%
\bibitem [{\citenamefont {Kawasaki}\ \emph {et~al.}(2022)\citenamefont
  {Kawasaki}, \citenamefont {Takase}, \citenamefont {Nomura}, \citenamefont
  {Miki}, \citenamefont {Terai}, \citenamefont {Yabuno}, \citenamefont {China},
  \citenamefont {Asavanant}, \citenamefont {Endo}, \citenamefont {Yoshikawa},\
  and\ \citenamefont {Furusawa}}]{Kawasaki2022}%
  \BibitemOpen
  \bibfield  {author} {\bibinfo {author} {\bibfnamefont {A.}~\bibnamefont
  {Kawasaki}}, \bibinfo {author} {\bibfnamefont {K.}~\bibnamefont {Takase}},
  \bibinfo {author} {\bibfnamefont {T.}~\bibnamefont {Nomura}}, \bibinfo
  {author} {\bibfnamefont {S.}~\bibnamefont {Miki}}, \bibinfo {author}
  {\bibfnamefont {H.}~\bibnamefont {Terai}}, \bibinfo {author} {\bibfnamefont
  {M.}~\bibnamefont {Yabuno}}, \bibinfo {author} {\bibfnamefont
  {F.}~\bibnamefont {China}}, \bibinfo {author} {\bibfnamefont
  {W.}~\bibnamefont {Asavanant}}, \bibinfo {author} {\bibfnamefont
  {M.}~\bibnamefont {Endo}}, \bibinfo {author} {\bibfnamefont {J.-i.}\
  \bibnamefont {Yoshikawa}},\ and\ \bibinfo {author} {\bibfnamefont
  {A.}~\bibnamefont {Furusawa}},\ }\bibfield  {title} {\bibinfo {title}
  {Generation of highly pure single-photon state at telecommunication
  wavelength},\ }\href {https://doi.org/10.1364/OE.460583} {\bibfield
  {journal} {\bibinfo  {journal} {Opt. Expr.}\ }\textbf {\bibinfo {volume}
  {30}},\ \bibinfo {pages} {24831} (\bibinfo {year} {2022})}\BibitemShut
  {NoStop}%
\bibitem [{\citenamefont {Takase}\ \emph {et~al.}(2022)\citenamefont {Takase},
  \citenamefont {Kawasaki}, \citenamefont {Jeong}, \citenamefont {Endo},
  \citenamefont {Kashiwazaki}, \citenamefont {Kazama}, \citenamefont {Enbutsu},
  \citenamefont {Watanabe}, \citenamefont {Umeki}, \citenamefont {Miki},
  \citenamefont {Terai}, \citenamefont {Yabuno}, \citenamefont {China},
  \citenamefont {Asavanant}, \citenamefont {Yoshikawa},\ and\ \citenamefont
  {Furusawa}}]{Takase2022}%
  \BibitemOpen
  \bibfield  {author} {\bibinfo {author} {\bibfnamefont {K.}~\bibnamefont
  {Takase}}, \bibinfo {author} {\bibfnamefont {A.}~\bibnamefont {Kawasaki}},
  \bibinfo {author} {\bibfnamefont {B.~K.}\ \bibnamefont {Jeong}}, \bibinfo
  {author} {\bibfnamefont {M.}~\bibnamefont {Endo}}, \bibinfo {author}
  {\bibfnamefont {T.}~\bibnamefont {Kashiwazaki}}, \bibinfo {author}
  {\bibfnamefont {T.}~\bibnamefont {Kazama}}, \bibinfo {author} {\bibfnamefont
  {K.}~\bibnamefont {Enbutsu}}, \bibinfo {author} {\bibfnamefont
  {K.}~\bibnamefont {Watanabe}}, \bibinfo {author} {\bibfnamefont
  {T.}~\bibnamefont {Umeki}}, \bibinfo {author} {\bibfnamefont
  {S.}~\bibnamefont {Miki}}, \bibinfo {author} {\bibfnamefont {H.}~\bibnamefont
  {Terai}}, \bibinfo {author} {\bibfnamefont {M.}~\bibnamefont {Yabuno}},
  \bibinfo {author} {\bibfnamefont {F.}~\bibnamefont {China}}, \bibinfo
  {author} {\bibfnamefont {W.}~\bibnamefont {Asavanant}}, \bibinfo {author}
  {\bibfnamefont {J.-i.}\ \bibnamefont {Yoshikawa}},\ and\ \bibinfo {author}
  {\bibfnamefont {A.}~\bibnamefont {Furusawa}},\ }\bibfield  {title} {\bibinfo
  {title} {Generation of {{S}}chr{\"o}dinger cat states with {{Wigner}}
  negativity using continuous-wave low-loss waveguide optical parametric
  amplifier},\ }\href {https://doi.org/10.1364/OE.454123} {\bibfield  {journal}
  {\bibinfo  {journal} {Opt. Expr.}\ }\textbf {\bibinfo {volume} {30}},\
  \bibinfo {pages} {14161} (\bibinfo {year} {2022})}\BibitemShut {NoStop}%
\bibitem [{\citenamefont {Guo}\ \emph {et~al.}(2020)\citenamefont {Guo},
  \citenamefont {Breum}, \citenamefont {Borregaard}, \citenamefont {Izumi},
  \citenamefont {Larsen}, \citenamefont {Gehring}, \citenamefont {Christandl},
  \citenamefont {{Neergaard-Nielsen}},\ and\ \citenamefont
  {Andersen}}]{Guo2020}%
  \BibitemOpen
  \bibfield  {author} {\bibinfo {author} {\bibfnamefont {X.}~\bibnamefont
  {Guo}}, \bibinfo {author} {\bibfnamefont {C.~R.}\ \bibnamefont {Breum}},
  \bibinfo {author} {\bibfnamefont {J.}~\bibnamefont {Borregaard}}, \bibinfo
  {author} {\bibfnamefont {S.}~\bibnamefont {Izumi}}, \bibinfo {author}
  {\bibfnamefont {M.~V.}\ \bibnamefont {Larsen}}, \bibinfo {author}
  {\bibfnamefont {T.}~\bibnamefont {Gehring}}, \bibinfo {author} {\bibfnamefont
  {M.}~\bibnamefont {Christandl}}, \bibinfo {author} {\bibfnamefont {J.~S.}\
  \bibnamefont {{Neergaard-Nielsen}}},\ and\ \bibinfo {author} {\bibfnamefont
  {U.~L.}\ \bibnamefont {Andersen}},\ }\bibfield  {title} {\bibinfo {title}
  {Distributed quantum sensing in a continuous-variable entangled network},\
  }\href {https://doi.org/10.1038/s41567-019-0743-x} {\bibfield  {journal}
  {\bibinfo  {journal} {Nat. Phys.}\ }\textbf {\bibinfo {volume} {16}},\
  \bibinfo {pages} {281} (\bibinfo {year} {2020})}\BibitemShut {NoStop}%
\bibitem [{\citenamefont {Miki}\ \emph {et~al.}(2013)\citenamefont {Miki},
  \citenamefont {Yamashita}, \citenamefont {Terai},\ and\ \citenamefont
  {Wang}}]{Miki2013}%
  \BibitemOpen
  \bibfield  {author} {\bibinfo {author} {\bibfnamefont {S.}~\bibnamefont
  {Miki}}, \bibinfo {author} {\bibfnamefont {T.}~\bibnamefont {Yamashita}},
  \bibinfo {author} {\bibfnamefont {H.}~\bibnamefont {Terai}},\ and\ \bibinfo
  {author} {\bibfnamefont {Z.}~\bibnamefont {Wang}},\ }\bibfield  {title}
  {\bibinfo {title} {High performance fiber-coupled {{NbTiN}} superconducting
  nanowire single photon detectors with {{Gifford-McMahon}} cryocooler.},\
  }\href {https://doi.org/10.1364/OE.21.010208} {\bibfield  {journal} {\bibinfo
   {journal} {Opt. Expr.}\ }\textbf {\bibinfo {volume} {21}},\ \bibinfo {pages}
  {10208} (\bibinfo {year} {2013})}\BibitemShut {NoStop}%
\bibitem [{\citenamefont {Yamashita}\ \emph {et~al.}(2013)\citenamefont
  {Yamashita}, \citenamefont {Miki}, \citenamefont {Terai},\ and\ \citenamefont
  {Wang}}]{Yamashita2013}%
  \BibitemOpen
  \bibfield  {author} {\bibinfo {author} {\bibfnamefont {T.}~\bibnamefont
  {Yamashita}}, \bibinfo {author} {\bibfnamefont {S.}~\bibnamefont {Miki}},
  \bibinfo {author} {\bibfnamefont {H.}~\bibnamefont {Terai}},\ and\ \bibinfo
  {author} {\bibfnamefont {Z.}~\bibnamefont {Wang}},\ }\bibfield  {title}
  {\bibinfo {title} {Low-filling-factor superconducting single photon detector
  with high system detection efficiency},\ }\href
  {https://doi.org/10.1364/oe.21.027177} {\bibfield  {journal} {\bibinfo
  {journal} {Opt. Expr.}\ }\textbf {\bibinfo {volume} {21}},\ \bibinfo {pages}
  {27177} (\bibinfo {year} {2013})}\BibitemShut {NoStop}%
\bibitem [{\citenamefont {Neuhaus}\ \emph {et~al.}(2024)\citenamefont
  {Neuhaus}, \citenamefont {Croquette}, \citenamefont {Metzdorff},
  \citenamefont {Chua}, \citenamefont {Jacquet}, \citenamefont {Journeaux},
  \citenamefont {Heidmann}, \citenamefont {Briant}, \citenamefont {Jacqmin},
  \citenamefont {Cohadon},\ and\ \citenamefont {Del{\'e}glise}}]{Neuhaus2024}%
  \BibitemOpen
  \bibfield  {author} {\bibinfo {author} {\bibfnamefont {L.}~\bibnamefont
  {Neuhaus}}, \bibinfo {author} {\bibfnamefont {M.}~\bibnamefont {Croquette}},
  \bibinfo {author} {\bibfnamefont {R.}~\bibnamefont {Metzdorff}}, \bibinfo
  {author} {\bibfnamefont {S.}~\bibnamefont {Chua}}, \bibinfo {author}
  {\bibfnamefont {P.-E.}\ \bibnamefont {Jacquet}}, \bibinfo {author}
  {\bibfnamefont {A.}~\bibnamefont {Journeaux}}, \bibinfo {author}
  {\bibfnamefont {A.}~\bibnamefont {Heidmann}}, \bibinfo {author}
  {\bibfnamefont {T.}~\bibnamefont {Briant}}, \bibinfo {author} {\bibfnamefont
  {T.}~\bibnamefont {Jacqmin}}, \bibinfo {author} {\bibfnamefont {P.-F.}\
  \bibnamefont {Cohadon}},\ and\ \bibinfo {author} {\bibfnamefont
  {S.}~\bibnamefont {Del{\'e}glise}},\ }\bibfield  {title} {\bibinfo {title}
  {Python {{Red Pitaya Lockbox}} ({{PyRPL}}): {{An}} open source software
  package for digital feedback control in quantum optics experiments},\ }\href
  {https://doi.org/10.1063/5.0178481} {\bibfield  {journal} {\bibinfo
  {journal} {Review of Scientific Instruments}\ }\textbf {\bibinfo {volume}
  {95}},\ \bibinfo {pages} {033003} (\bibinfo {year} {2024})}\BibitemShut
  {NoStop}%
\bibitem [{\citenamefont {Appel}\ \emph {et~al.}(2007)\citenamefont {Appel},
  \citenamefont {Hoffman}, \citenamefont {Figueroa},\ and\ \citenamefont
  {Lvovsky}}]{appel_electronic_2007}%
  \BibitemOpen
  \bibfield  {author} {\bibinfo {author} {\bibfnamefont {J.}~\bibnamefont
  {Appel}}, \bibinfo {author} {\bibfnamefont {D.}~\bibnamefont {Hoffman}},
  \bibinfo {author} {\bibfnamefont {E.}~\bibnamefont {Figueroa}},\ and\
  \bibinfo {author} {\bibfnamefont {A.~I.}\ \bibnamefont {Lvovsky}},\
  }\bibfield  {title} {\bibinfo {title} {Electronic noise in optical homodyne
  tomography},\ }\href {https://doi.org/10.1103/PhysRevA.75.035802} {\bibfield
  {journal} {\bibinfo  {journal} {Phys. Rev. A}\ }\textbf {\bibinfo {volume}
  {75}},\ \bibinfo {pages} {35802} (\bibinfo {year} {2007})}\BibitemShut
  {NoStop}%
\bibitem [{\citenamefont {{Neergaard-Nielsen}}(2008)}]{Neergaard-Nielsen2008}%
  \BibitemOpen
  \bibfield  {author} {\bibinfo {author} {\bibfnamefont {J.~S.}\ \bibnamefont
  {{Neergaard-Nielsen}}},\ }\emph {\bibinfo {title} {Generation of Single
  Photons and {{Schr{\"o}dinger}} Kitten States of Light}},\ \href
  {https://nbi.ku.dk/english/theses/phd-theses/jonas-schou-neergaard-nielsen/jonas_phd.pdf}
  {Ph.D. thesis},\ \bibinfo  {school} {University of Copenhagen} (\bibinfo
  {year} {2008})\BibitemShut {NoStop}%
\bibitem [{\citenamefont {Lvovsky}\ and\ \citenamefont
  {Raymer}(2009)}]{Lvovsky2009a}%
  \BibitemOpen
  \bibfield  {author} {\bibinfo {author} {\bibfnamefont {A.~I.}\ \bibnamefont
  {Lvovsky}}\ and\ \bibinfo {author} {\bibfnamefont {M.~G.}\ \bibnamefont
  {Raymer}},\ }\bibfield  {title} {\bibinfo {title} {Continuous-variable
  optical quantum state tomography},\ }\href
  {https://doi.org/10.1103/RevModPhys.81.299} {\bibfield  {journal} {\bibinfo
  {journal} {Rev. Mod. Phys.}\ }\textbf {\bibinfo {volume} {81}},\ \bibinfo
  {pages} {299} (\bibinfo {year} {2009})}\BibitemShut {NoStop}%
\end{thebibliography}%

\end{document}